%% file: main.tex
\documentclass[%
reprint,
superscriptaddress,
nofootinbib,
amsmath,amssymb,
aps,
prl,
]{revtex4-1}
\usepackage[english]{babel}
\usepackage{blindtext}
\usepackage[utf8]{inputenc}
\usepackage{graphicx}
\usepackage{overpic}
 \usepackage{rotating}
 \usepackage{mwe, tikz}
 \usepackage{amsmath,amssymb}
\usepackage{pgfplots}
\usetikzlibrary{calc,arrows,decorations.markings,shapes.geometric} 
\usepackage{amssymb}
\usepackage[inline]{trackchanges}
\usepackage{lipsum}
\usepackage{romannum}
\usepackage{float}
\usepackage{romannum}
\usepackage{perpage}
\MakePerPage{footnote}

\usepackage{xfp}
\usepackage{graphicx}
\usepackage{amsmath}
\usepackage{mathtools} 
\DeclarePairedDelimiter\norm{\lVert}{\rVert}
\usepackage{color}
\usepackage{overpic}
\usepackage[export]{adjustbox}
\usepackage{mwe, tikz}
\usepackage{rotating}
\usepackage{latexsym}
\usepackage{csquotes}
\usepackage{tikz} 
\usetikzlibrary{calc,arrows,decorations.markings} 
\usepackage{pict2e}
\usepackage{pgfplots}
\usepackage{float}
\usepackage{hyperref}
\usepackage{xr}
\usepackage{rotating}
\usepackage{amsmath,amssymb}
\usepackage{fp}
\usetikzlibrary{calc} 
\usepackage{booktabs, tabularx}
\usepackage{romannum}
\usepackage{bm}
\usepackage{float}
\usepackage{adjustbox}
\usepackage{booktabs, tabularx}
\usepackage{longtable}
\usepackage{romannum}
\usepackage{footnote}
\usepackage{hyperref}
\usepackage[symbol*]{footmisc}

\input{macros}
\soulregister\ref7
\soulregister\cite7
\soulregister\citeA7
\soulregister\romannum7
\soulregister\section7

\definecolor{darkolivegreen}{rgb}{0.33, 0.42, 0.18}
\definecolor{darkspringgreen}{rgb}{0.09, 0.45, 0.27}
\definecolor{darkslategray}{rgb}{0.18, 0.31, 0.31}
\definecolor{darkred}{rgb}{0.55, 0.0, 0.0}

\addeditor{AN}
\pgfplotsset{compat=1.18}
\begin{document}
\pagenumbering{arabic}

\title{Neural Network Interatomic Potentials
For Open Surface Nano-mechanics Applications
}

\author{Amirhossein D. Naghdi$^{*}$}
\affiliation{%
NOMATEN Centre of Excellence, National Center for 
 Nuclear Research, ul. A. Sołtana 7, 05-400 Swierk/Otwock
}%
\affiliation{%
IDEAS NCBR, ul. Chmielna 69, 00-801, Warsaw, Poland
}%
\author{Franco Pellegrini}
\affiliation{%
 International School for Advanced Studies (SISSA), 
 Via Bonomea, 265, I-34136 Trieste, Italy
}%

\author{Emine K\"uc\"ukbenli}
\affiliation{%
Nvidia Corporation, Santa Clara, CA, USA}
\affiliation{%
John A. Paulson School of Engineering and Applied Sciences, Harvard University, Cambridge, Massachusetts 02138, USA}

\author{Dario Massa}
\affiliation{%
NOMATEN Centre of Excellence, National Center for 
 Nuclear Research, ul. A. Sołtana 7, 05-400 Swierk/Otwock
}%
\affiliation{%
IDEAS NCBR, ul. Chmielna 69, 00-801, Warsaw, Poland
}%
\author{F. Javier Dominguez--Gutierrez}
\affiliation{%
NOMATEN Centre of Excellence, National Center for 
 Nuclear Research, ul. A. Sołtana 7, 05-400 Swierk/Otwock
}%

\author{Efthimios Kaxiras}
\affiliation{%
John A. Paulson School of Engineering and Applied Sciences, Harvard University, Cambridge, Massachusetts 02138, USA}
\affiliation{%
Department of Physics, Harvard University, Cambridge, Massachusetts 02138, USA}

\author{Stefanos Papanikolaou$^{*}$}
\affiliation{%
NOMATEN Centre of Excellence, National Center for 
 Nuclear Research, ul. A. Sołtana 7, 05-400 Swierk/Otwock
}%

\def\thefootnote{$*$}
\footnotetext{Corresponding authors\\N.D.A., E-mail: \url{Amirhossein.Naghdi@ncbj.gov.pl}\\S.P., E-mail: \url{Stefanos.Papanikolaou@ncbj.gov.pl}}
\def\thefootnote{\arabic{footnote}}

\begin{abstract}

Material characterization in nano-mechanical tests requires precise interatomic potentials for the computation of
atomic energies and forces with near-quantum accuracy. For such purposes, we develop a robust neural-network interatomic potential (NNIP), and we provide a test for the example of molecular dynamics (MD) nanoindentation, and
the case of body-centered cubic crystalline molybdenum (Mo). We employ a similarity measurement protocol, using
standard local environment descriptors, to select ab initio configurations for the training dataset that capture the
behavior of the indented sample. We find that it is critical to include generalized stacking fault (GSF) configurations,
featuring a dumbbell interstitial on the surface, to capture dislocation cores, and also high-temperature configurations with frozen atom layers for the indenter tip contact. We develop a NNIP with distinct dislocation nucleation
mechanisms, realistic generalized stacking fault energy (GSFE) curves, and an informative energy landscape for the
atoms on the sample surface during nanoindentation. We compare our NNIP results with nanoindentation simulations,
performed with three existing potentials—an embedded atom method (EAM) potential, a gaussian approximation potential (GAP), and a tabulated GAP (tabGAP) potential—that predict different dislocation nucleation mechanisms,
and display the absence of essential information on the shear stress at the sample surface in the elastic region. We
believe that these features render specialized NNIPs essential for simulations of nanoindentation and nano-mechanics
with near-quantum accuracy.

\end{abstract}
\maketitle

\section{\label{sec:introduction}Introduction}

\begin{figure*}[ht]
  \centering
  \includegraphics[width=0.98\textwidth]{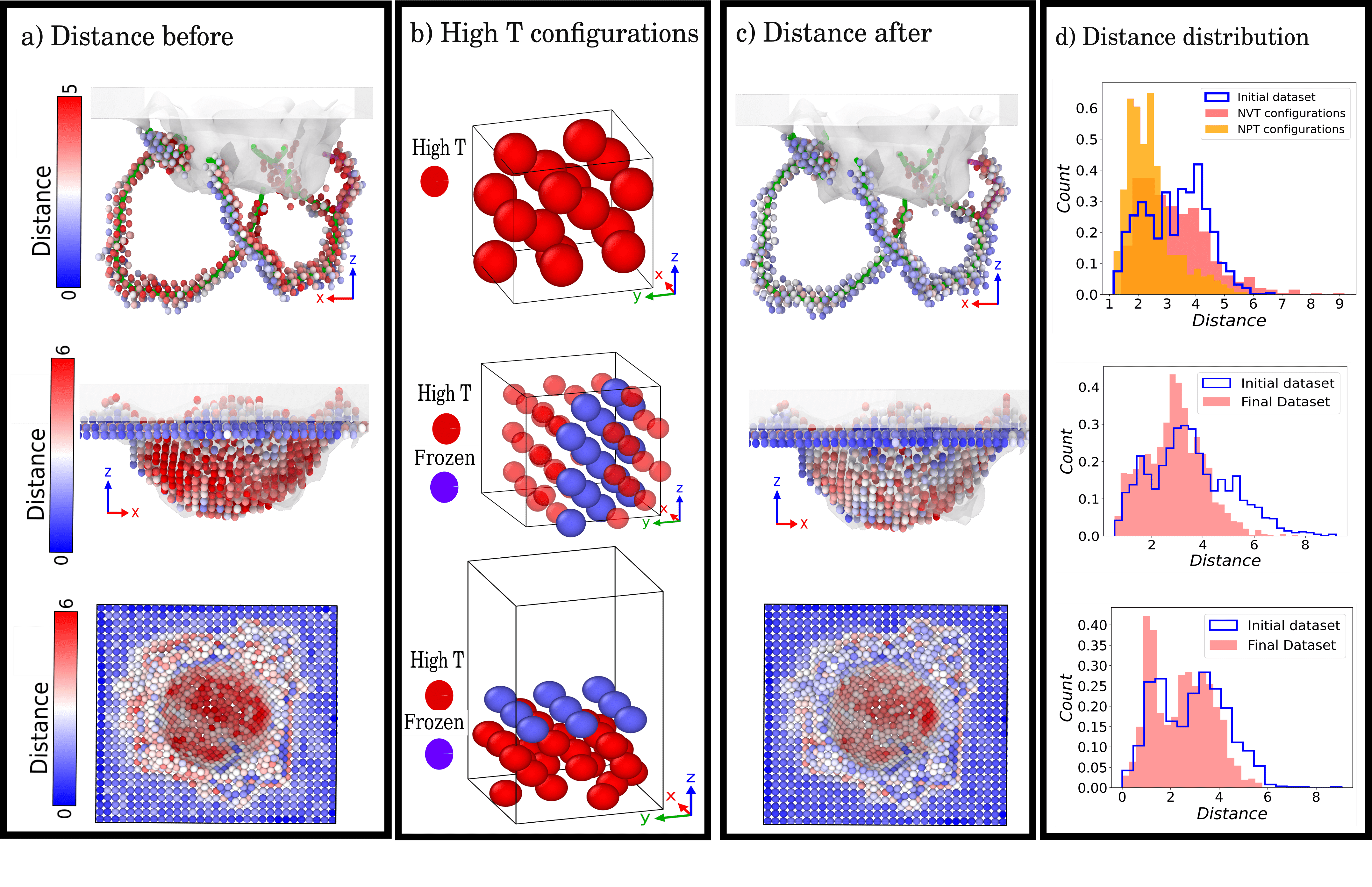}
  \caption{Illustration of the novel configurations discovered in this study, which correspond to distinct regions of an indented sample. The figures are arranged horizontally to demonstrate the correlation between them. a) The distances of the dislocation cores, atoms beneath the indenter tip, and pileup atoms on the surface from the original dataset. b) The newly detected high-temperature configurations, which are associated with diverse regions of the indented sample, in terms of their distance. c) The distance of the aforementioned regions shown in panel (a) after incorporating the newly found configurations to the dataset. d) The distribution of distances of the aforementioned atoms from the dataset before and after incorporating the newly introduced configurations.} 
  \label{fig:newconfigs}
\end{figure*}

Nano-mechanical tests serve as essential tools for 
probing the mechanical properties of materials at the
nanoscale. Techniques such as nano-tensile/compression \cite{KIM20095245,KIM201246, amirtensile}, nanoindentation ~\cite{SCHUH200632,VARILLAS2017431,KURPASKA2022110639,PATHAK20151,cryst7100321, REMINGTON2014378, GAGEL2016399, naghdi2022dynamic}, and creep testing ~\cite{Taneike2003}
play a pivotal role in revealing the intrinsic properties
of materials. This understanding, in turn, facilitates the
design and production of innovative materials capable of
functioning in extreme environments. These tests involve
subjecting the material to controlled strain/stress at the
nanoscale, enabling researchers to gain valuable insights
into its mechanical response. This knowledge is crucial in
the field of defect physics, as nano-mechanical tests 
provide a means to investigate the mechanisms of defects 
nucleation and their impact on the mechanical performance
of materials under extreme conditions. In this study, we
aim to present a comprehensive method for simulating
nano-mechanical tests, taking nanoindentation as an
example, on crystalline materials using neural-network interatomic potentials (NNIPs).

Nano-mechanical test techniques find application in
a several areas of materials science.
Specifically, in situ techniques ~\cite{Haque2002, DeHosson2006} contribute significantly
to the understanding of material deformation under 
controlled applied stress or strain, while the specimen
is simultaneously observed/measured by electron microscopic
devices. These methodologies play a pivotal role
in exploring materials properties at the nano scale, offering
insights into the intrinsic properties of materials, such
as the strength of each crystalline grain. Furthermore, these
techniques prove invaluable in investigating temperature-
related deformation mechanisms inherent in crystalline
materials. The focus of this paper is on nanoindentation
testing, a widely utilized method for assessing material
properties on open surfaces. This technique yields results
for various properties, encompassing hardness, strength,
dislocation nucleation mechanisms, dislocation density,
grain boundary effects, and dislocation junction formations
~\cite{durst, MaierKiener2017, CASALS200755, PhysRevB.67.245405, SWADDIWUDHIPONG20061117, DOMINGUEZGUTIERREZ2021141912, article, PhysRevB.76.165422, VOYIADJIS2012205, plummer}. However, it is essential to note that
nanoindentation testing involves intricate defect nucleation
mechanisms and plastic deformations, rendering accurate
modeling a formidable challenge within the realm of 
computational materials science.

Various computational methods, such as finite element
methods (FEM) ~\cite{FRYDRYCH2023104644, 10.1063/1.369178, LICHINCHI1998240}, discrete dislocation dynamics
(DDD) ~\cite{ZHOU20101565, MOTZ20081942, SONG2019332}, and molecular dynamics (MD) ~\cite{XU2024112733, PhysRevMaterials.7.043603, PhysRevB.107.094109, karimi2023serrated, karimi2023tuning}, 
are employed for modeling nano-mechanical testing.
In FEM, numerical solutions to differential equations in
mathematical models are used to approximate and analyze
the complex behavior of materials. While FEM is
useful in certain cases, it lacks atomic-level information
and, therefore, does not achieve quantum accuracy. 
DDD simulations explicitly describe dislocation
dynamics at the mesoscale, but are limited in predicting
defect properties with quantum-level accuracy. In contrast,
MD simulations can provide atomic-level insights
into the dislocation dynamics of materials, given the use
of interatomic potentials finely tuned for nano-mechanics
in the simulations.

Machine-learned force fields (MLFFs) ~\cite{PhysRevLett.98.146401, PhysRevLett.104.136403, THOMPSON2015316, 15M1054183, PhysRevB.97.184307, PhysRevLett.120.143001, doi:10.1063/1.4966192, Shaidu2021, LI2021100359} offer 
a reliable means for modeling nano-mechanical tests with
quantum-level accuracy. Various MLFF types, such as
Gaussian Approximation Potentials (GAP) ~\cite{PhysRevLett.104.136403} and its
tabulated version (tabGAP) ~\cite{PhysRevMaterials.6.083801}, as well as active learning 
methods like FLARE ~\cite{Vandermause2020}, are available in the literature. In addition,
NNIPs exhibit exceptional accuracy in
predicting atomic energies and forces ~\cite{pellegrini2023panna, Musaelian2023, Batatia2022Design}, 
overcoming the time and system
size limitations inherent in traditional ab initio molecular
dynamics (AIMD) simulations. Given the ability of NNIPs 
to learn complex functions, such as the energy landscape
of an extended dislocation in a metallic crystal, 
they prove to be excellent tools for modeling
nano-mechanical testing simulations. MLFFs have
been successfully applied to various problems, including
catalysis ~\cite{owen2023stability, owen2023unraveling}, point defects modeling ~\cite{PhysRevMaterials.5.103803, PhysRevB.100.144105}, multi-
component materials modeling \cite{Nikoulis_2021, PhysRevB.104.104101}, and multi-phase
systems ~\cite{Shaidu2021}, demonstrating their versatility. However,
the exploration of nano-mechanical testing simulations
using MLFFs is an area that remains to be fully explored.

\begin{figure*}[ht]
  \centering
  \includegraphics[width=0.98\textwidth]{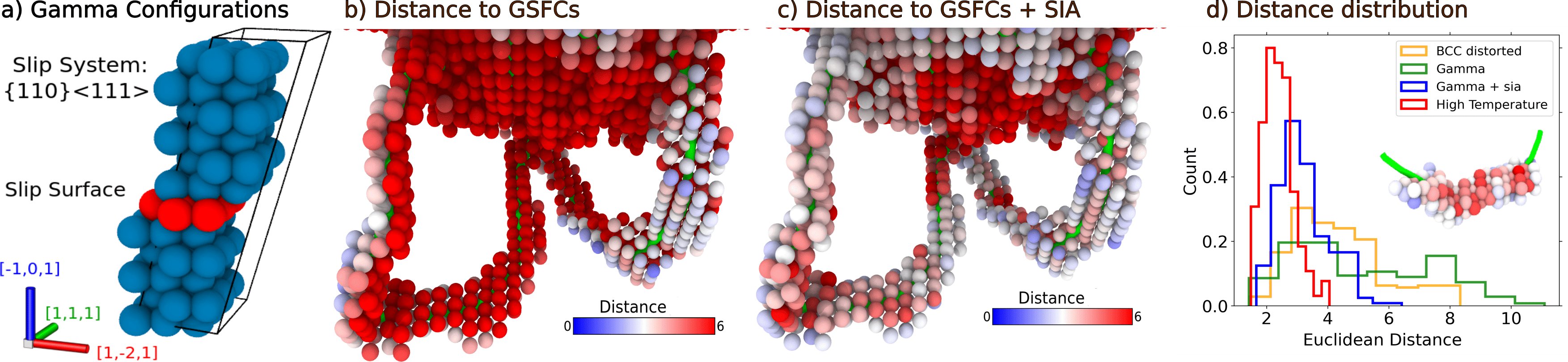}
  \caption{The impact of incorporating GSFCs on the distances of atoms on dislocation cores from the dataset. a) A schematic representation of the GSFCs integrated into the dataset. b) Addition of GSFCs reduces the distances of atoms on the slip plane of dislocation cores from the dataset. c) By including a SIA on the surface of GSFCs, all atoms on the dislocation cores can be covered. d) The distribution of atom distances on dislocation cores reveals that GSFCs with SIA can effectively cover dislocation cores.} 
  \label{fig:GSFCs}
\end{figure*}

In this paper, we present a study focused on the 
development of a quantum-level-accurate NNIP by 
enhancing a starting dataset sourced from the literature ~\cite{dataset}
within the PANNA (Properties from Artificial Neural
Network Architectures) framework ~\cite{panna}. We compare the
Behler-Parrinello descriptor vectors ~\cite{PhysRevLett.98.146401} of the training
dataset with those of a single crystal BCC Molybdenum
configuration, indented with an embedded atom model
(EAM) potential ~\cite{ESalonen_2003}, to determine how closely the 
training dataset resembles the indentation process. To improve 
the accuracy of the potential, we introduce high
temperature configurations with a frozen layer and 
generalized stacking fault (GSF) configurations with an 
interstitial on the gamma surface. These configurations are
designed to closely mimic atoms in the dislocation cores,
on the surface, and under the indenter tip. Our results
show that including these configurations in the training
dataset reduces the distance between the atoms the potential
is trained on and the indented sample. Finally, we
present the results of an MD nanoindentation simulation
using the potential trained with the modified dataset.

\section{\label{sec:Results}Results}

\subsection{NNIP Dataset Explores the Intricacies of Every Atomic Landscape During Nanoindentation}

First, we applied a similarity measurement methodology, resulting in an ideal dataset for nano-mechanical applications, particularly in the context of nanoindentation simulations.
This method assesses similarity by comparing distances between atoms in two configurations/datasets, using modified Behler-Parinnelo descriptors ~\cite{C6SC05720A, PhysRevLett.98.146401}. Then it focuses on finding the "largest minimum" distance of atomic 
descriptors in two configurations/datasets (see Methods). We employed this method for comparing configuration types in a dataset from literature \cite{dataset} to atomic landscapes in an indented sample accuaired with empirical methods ~\cite{ESalonen_2003}, aiming to gain 
insights into the dataset's suitability for studying nanoindentation behavior and uncovering underlying mechanisms. 

To curate a well-designed dataset for the nanoindentation of single crystalline
Molybdenum (Mo), we identified unique configurations with a close distance to defected regions of an indented configuration. This encompasses point defects, situated near the sample surface, as well as extended defects like dislocation lines and junctions.
Fig.\ref{fig:newconfigs}(a) summarizes the three defected regions, and their color-coded distance to the initial dataset, of an indented sample. This includes the extended dislocation lines and junctions (top), point defects under the indenter tip (middle) and the surface atoms (bottom). 
An interatomic potential employed for nano-mechanical simulations must effectively 
capture the intricate energy landscape of defects that arise during the simulation. Consequently, we 
identified distinctive configurations to ensure comprehensive coverage of this complexity (Fig. 
\ref{fig:newconfigs}(b)). Our findings suggest that high-temperature NPT configurations, in contrast to NVT configurations, have a close distance to dislocation cores. Additionally, it was observed 
that these configurations address the defected regions beneath the indenter tip when a layer of atoms is kept frozen. The ultimate refinement of 
the dataset is optimized by introducing a void atop high-temperature configurations with the frozen layers, particularly benefiting the surface 
atoms.
The effect of aformentioned dataset modifications is demonstrated in Fig.\ref{fig:newconfigs}(c-d). Ultimately, the distances between atoms from the indented sample and the dataset are reduced, evident in the narrowing of the distance distribution width.

Moreover, it was also observed that Generalized 
Stacking Fault Configurations (GSFCs) 
incorporating a dumbbell self-interstitial atom (SIA) on the top have a close distance to atoms within the dislocation cores. 
A representative GSFC is illustrated in Fig.\ref{fig:GSFCs}(a). A GSFC without any point 
defects would solely exhibit proximity to the atoms on the plane where the dislocation is 
slipping, which is shown in Fig.\ref{fig:GSFCs}(b). GSFCs with a dumbbell interstitial positioned 
near the $\gamma$-surface (the slip surface), on 
the other hand, were found to be in close 
proximity to the dislocation cores (Fig.\ref{fig:GSFCs}(c)). This outcome is further 
evident in the distance distributions depicted in 
Fig. \ref{fig:GSFCs}(d). Distorted BCC and GSF configurations exhibit a tail surpassing the 
distance threshold, whereas high NPT high-temperature configurations along with GSFCs + SIA display a smaller tail. However, the high-temperature configurations were excluded from the 
training set due to their substantial energy variation, introducing challenges 
in effectively training a NNIP (Fig.\ref{fig:s1}(b)). GSFCs, on the other hand happen to have a smaller energy variation, which suggests a more accurate prediction for dislocation cores and also is closer to the concept of a defect in crystalls. GSFCs, conversely, exhibit a significantly smaller energy variation. This implies a more precise prediction for dislocation cores and aligns more closely with the concept of defects in crystals. Ultimately, the dataset for the NNIP is meticulously crafted, making it ideal for nanoindentation simulations. For a more detailed description of the distance metric and the configurations considered, please see Methods.

\subsection{NNIP Validations}

\subsubsection{NNIP Achieves Quantum-Accurate Predictions for Energies and Forces}

\begin{figure}[h!]
  \centering
  \includegraphics[width=0.47\textwidth]{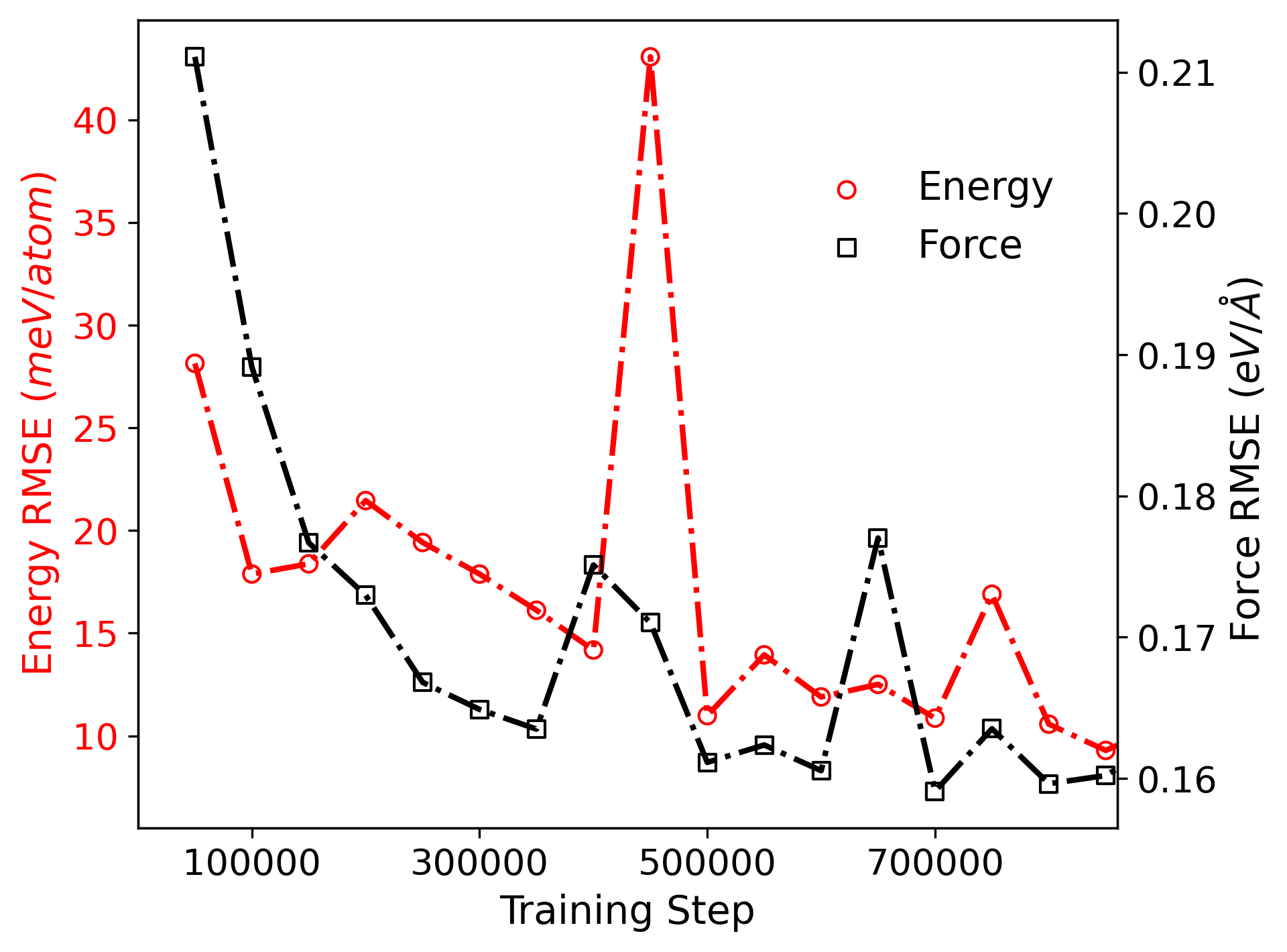}
  \caption{Prediction of the Energies and forces for the validation set during the NNIP training.} 
  \label{fig:val}
\end{figure}

We assessed the accuracy of the trained NNIP by calculating the root mean square error (RMSE) for both energies per atom (E-RMSE) and forces components (F-RMSE) at each checkpoint saved during training. To ensure the reliability of the final model on unseen data, $10 \%$ of configurations of each structure type were reserved for validation prior to training. Fig.~\ref{fig:val} shows that both the F-RMSE and E-RMSE decrease gradually as the network processes more data, reaching a plateau after 850K training steps with minimum values of 9.2 meV/atom and 0.16 eV/\AA, respectively.

\begin{figure}[h!]
  \centering
  \includegraphics[width=0.45\textwidth]{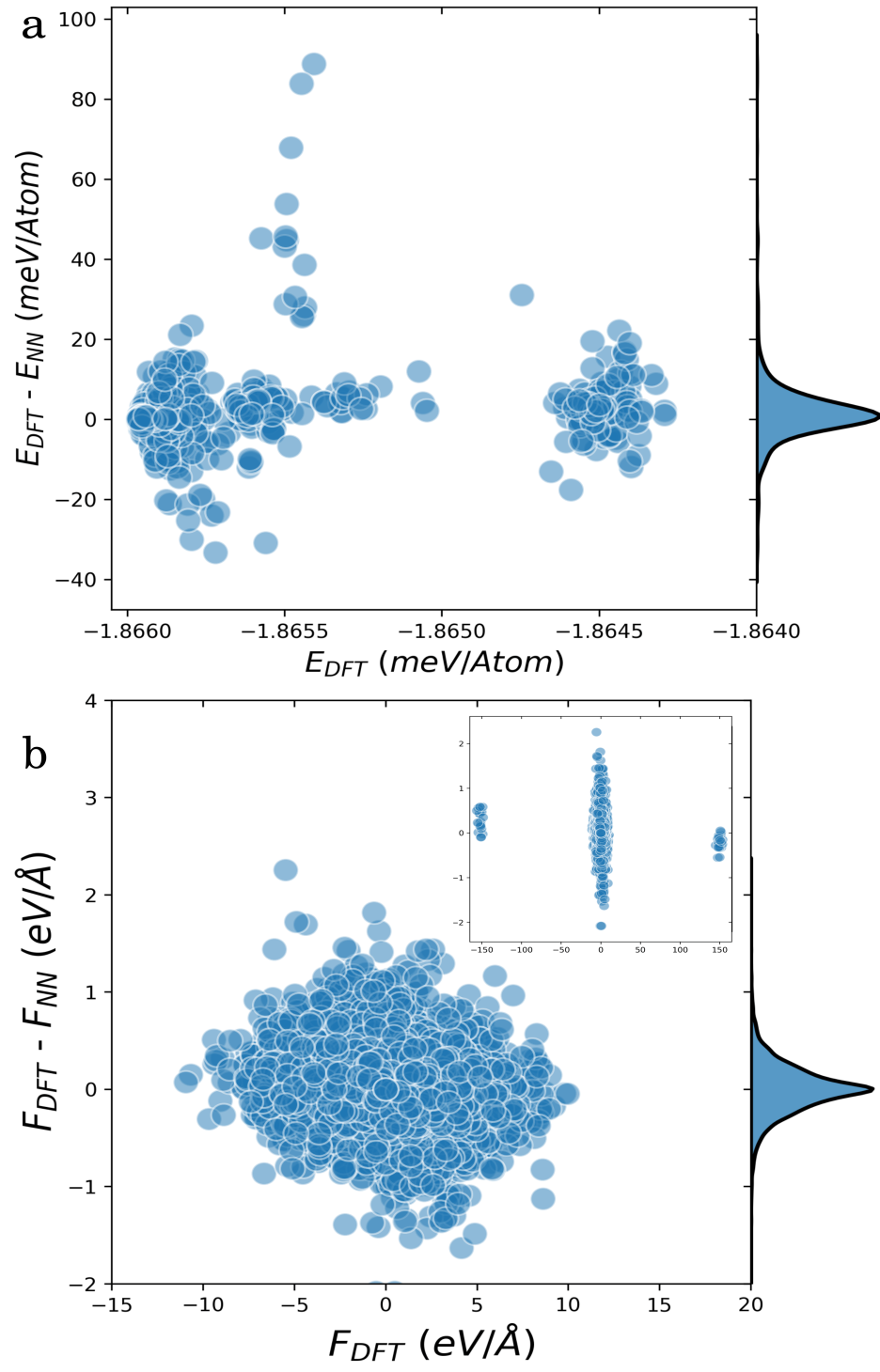}
  \caption{NNIP error on \textbf{(a)} total energies and \textbf{(b)} forces of each atom.} 
  \label{fig:EVal}
\end{figure}

The error distribution for both energies and forces are shown in the histogram plot of Fig.~\ref{fig:EVal}(a,b). The two islands in Fig.~\ref{fig:EVal}(a) are due to the energy difference between the pure and defected crystals. Also, the presence of three clusters in Fig.~\ref{fig:EVal}(b) is due to the large forces on the atoms in the defected configurations.

\begin{table*}[t!]
\centering
\resizebox{1.5\columnwidth}{!}{%
\begin{tabular}{crrrrrrr}
\hline
 & GAP & tabGap & EAM & NNIP	& DFT$^{a}$& DFT$^{b}$ & Exp$^{c}$\\
\hline
$C_{11}$ (GPa) & 478 (3.02\%) & 494 (6.47\%) & 465 (0.22\%) & 452 (2.59\%) & 459 & 468 & 464\\
$C_{12}$ (GPa) & 166 (4.40\%) & 146 (8.18\%) & 161 (1.26\%) &  121 (23.90\%) & 162 & 155 & 159\\
$C_{44}$ (GPa) & 108 (0.92\%) & 87 (20.18\%) & 109 (0\%) &  111 (1.83\%) & 97  & 100  &109 \\
$B$ (GPa) & 270 (8.00\%) & 262 (4.80\%) & 263 (5.20\%) &  231 (7.60\%) & 262 & - & 250\\
$\nu$ & 0.26 (10.34\%) & 0.23 (20.69\%) & 0.26 (10.34\%) &  0.21 (27.59\%) & 0.30 & - &0.29\\
\hline
\multicolumn{4}{l}{\small $^{a}$ This work.} \\
\multicolumn{4}{l}{\small $^{b}$ Reference \cite{dataset}}. \\
\multicolumn{4}{l}{\small $^{c}$ Reference \cite{rumble2019crc}.} \\
\end{tabular}
}
\caption{Elastic
constants $C_{ij}$, bulk modulus $B$, and Poisson ratio $\nu$, 
as obtained with the GAP, tabGAP, EAM/FS and the NNIP potentials compared to DFT done in this work and experimental data. In parenthesis is reported the modulus of the percentage error with respect to the experimental value.}
\label{tab:elastic}
\end{table*}

\subsubsection{Bulk Validation}

Next, we compare the elastic properties of the NNIP interatomic potential with both DFT and experimental results, as well as to those predicted by other interatomic potentials, such as GAP, tabGAP, and the EAM/FS potential, to evaluate the NNIP performance relative to other commonly used potentials. Table.~\ref{tab:elastic} summarizes the results of the comparison, explicitly reporting percentage errors with respect to the experimental values. The NNIP performs well for $C_{11}$, $C_{44}$ and $B$, with percentage errors below $8\%$ and in similar magnitude to GAP and EAM predictions.
We here stress that the accurate prediction of the shear modulus, $C_{44}$, is crucial for simulating the stresses that are applied to the surface of the sample during nanoindentation, and, following the good results of EAM and GAP for this measure, the NNIP proves itself to be promising for such applications. While the largest error for the NNIP concerns the prediction of $C_{12}$, it can still be considered within a reasonable range as it does not exceedingly influence the prediction on $B$\footnote{We here remind that $B=\dfrac{1}{3}(C_{11}+2C_{12})$.}.

\subsubsection{NNIP Accurately Predicts Generalized Stacking Fault Energies (GSFE)}

Finally, we compare the 
NNIP predictions for the GSFE against the DFT results, as well as other interatomic potentials mentioned in this work. The study focuses on the 
two most important slip systems of BCC crystals, namely the $\{110\}\langle\bar111\rangle$ and $\{121\}\langle\bar111\rangle$ families. The 
results were obtained for these directions in pure crystals. Additionally, since it was observed from Fig.~\ref{fig:GSFCs} that GSF $\{110\}\langle111\rangle$ configurations with a $
\langle111\rangle$ dumbbell interstitial on the surface are essential to cover the atomic environments of the 
dislocation core in terms of their distance to the indented samples, we also calculated and compared the GSFE for these configurations. 

\begin{figure}[h!]
  \centering
  \includegraphics[width=0.48\textwidth]{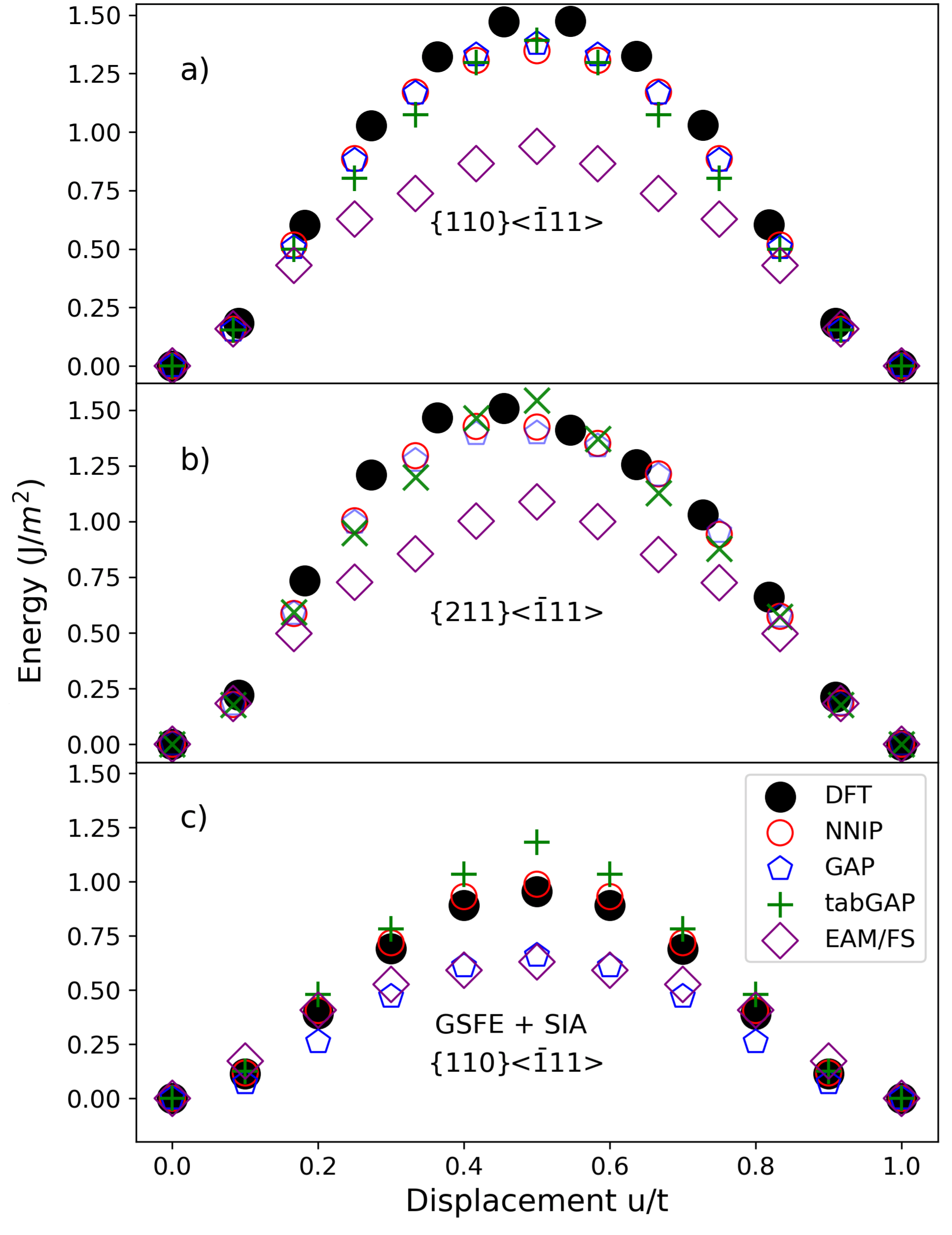}
  \caption{Generalized Stacking Fault Energy (GSFE) for single crystalline Molybdenum for: (a) $\{110\}\langle\bar111\rangle$ and (b) $\{211\}\langle\bar111\rangle$. 
  (c) The GSFE curve for the "GSFCs $+$ SIA" configurations.} 
  \label{fig:GSFE}
\end{figure}

Fig.~\ref{fig:GSFE} shows that all potentials 
predict the GSFE very accurately for both slip system families and pure crystals. However, EAM/FS displays errors of about $50\%$ and $32\%$ for the peak of the curve for $\{110\}\langle\bar111\rangle$ and $\{121\}\langle\bar111\rangle$ slip families, respectively. The configurations associated with these curves are crucial, as they represent the atoms on the slip plane of an indented sample, as illustrated and discussed in 
Fig.~\ref{fig:GSFCs}. 
While it is important for an interatomic potential to accurately predict the GSFE curve for reliable dislocation modeling, it is equally crucial for the potential to correctly predict the energies and forces on the atoms for 
the dislocation cores. Therefore, in addition to the GSFE curves for pure crystals, we calculated the GSFE curve for configurations with a $\langle111\rangle$ dumbbell interstitial on the surface. As depicted in Fig.~\ref{fig:GSFE}(c), NNIP is the potential that best predicts these energies, indicating the accurate simulation of dislocation dynamics during nanoindentation. In contrast, GAP and EAM potentials showed errors of $40\%$ and tabGAP showed an error of $20\%$ against DFT results, indicating their inability to accurately predict these values. This is discussed further in the following section.

\subsection{NNIP Nanoindentation predictions}

\begin{figure} 
    \centering
    \includegraphics[width=0.48\textwidth]{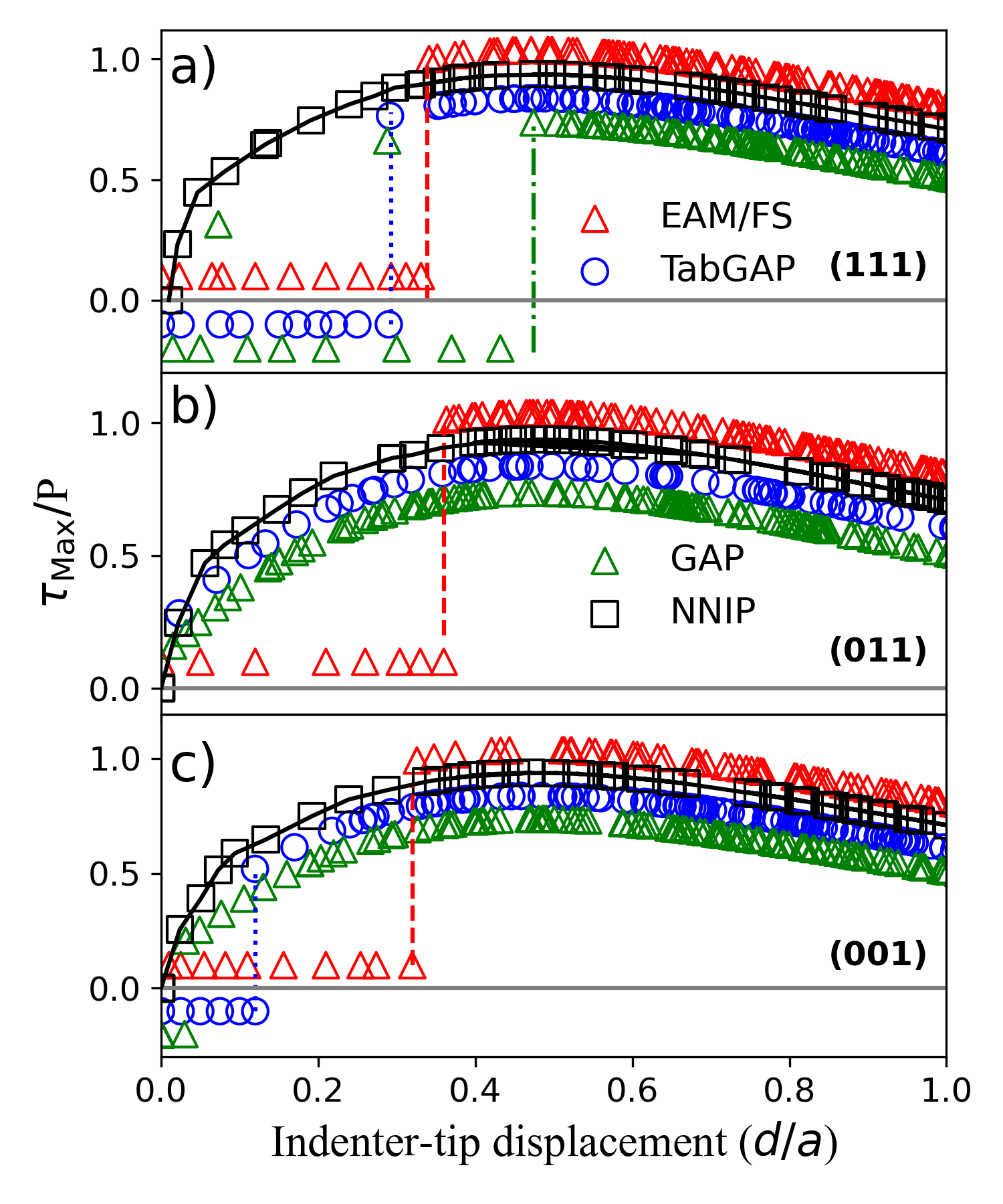}
    \caption{Hertzian calculation of normalized maximum shear
    stress by the applied pressure, $\tau_{\rm max}/P$, as 
    a function of normalized depth for main crystal orientations. 
    Surface information is needed in the interatomic potentials to 
    model nanoindentation induced plasticity in the 
    range of 0.0 to 0.475 $d/a$.
    To aid the  interpretation of the results, the values for EAM/FS were shifted by $-0.1$, the values for GAP were shifted by $+0.2$, and the values for tabGAP were shifted by $+0.1$ (all values in units of $\tau_{\rm max}/P$). 
    }
    \label{fig:maxstress}
\end{figure}
Plastic deformation 
does not initiate at the surface, 
but rather at a certain depth below it, 
typically a few atomic layers deep. 
This depth is known as the ``yield point'' or ``yield depth'', at which point the 
material begins to nucleate defects and dislocations under the 
applied load or stress. 
The initiation of the plastic deformation of the material 
occurs within the closest plastic region along the 
vertical $z$-axis beneath the spherical indenter tip.
In Fig.~\ref{fig:maxstress} we show results for the 
normalized maximum shear stress $\tau_{\rm max}/P$ 
which is a dimensionless quantity, with $P$ being
the applied pressure (Eq. 
\ref{eq:pressure}) and  $\tau_{\rm max}$ the shear stress (Eq. \ref{eq:shearMax}) calculated by using a linear elastic
contact mechanics formulation 
\cite{PhysRevMaterials.7.043603,VARILLAS2017431},
as a function of the displacement $d$ for [001], [011], and [111] 
main crystal orientations~\cite{PhysRevMaterials.7.043603}. A detailed explanation of the normalized shear stress calculation is provided in the Methods section. 

Our MD simulations report enough surface energy to model 
the nanoindentation induced plasticity as observed at distances 
close to the sample surface regardless of the crystal orientation, 
which is 
challenging for traditional and current ML
interatomic potentials for BCC Mo.
The modeling of the nanocontact of the indenter tip 
and the top atomic layers of the surface, from 0 to $\sim 0.3$ $d/a$ range with $d$ 
the indentation depth and $a$ the contact area, 
is important due to the nucleation of dislocation being dependent
on this mechanisms.\\

The GAP simulations provide valuable insights into the interaction
between the indenter tip and the top layer atoms for the (001) and (011) orientations.
However, for the (111) orientation, this information is lacking, resulting in a 
limitation in accurately modeling the nanoindentation test before the yield point.
This limitation arises due to the absence of the relevant atomic configurations in the training data for this
specific potential.
As a consequence, the tabGAP simulations follow a similar trend for the (111)
orientation, reflecting the lack of detailed information on the interaction between
the tip and the surface atoms.

In contrast, the NNIP simulations incorporate sufficient  information on
surface structures, allowing for a more accurate representation of the contact area. 
This is particularly important as the contact area depends on the applied load. 
The computed force between the tip and the atoms comprising the contact area is well-modeled in the NNIP simulations. 
The difference in spacing between data points in the elastic part of the graph is attributed to variations in the loading force, pressure, and maximum shear stress, which are considered in the contact area analysis.

\begin{figure}[b!]
    \centering
    \includegraphics[width=0.48\textwidth]{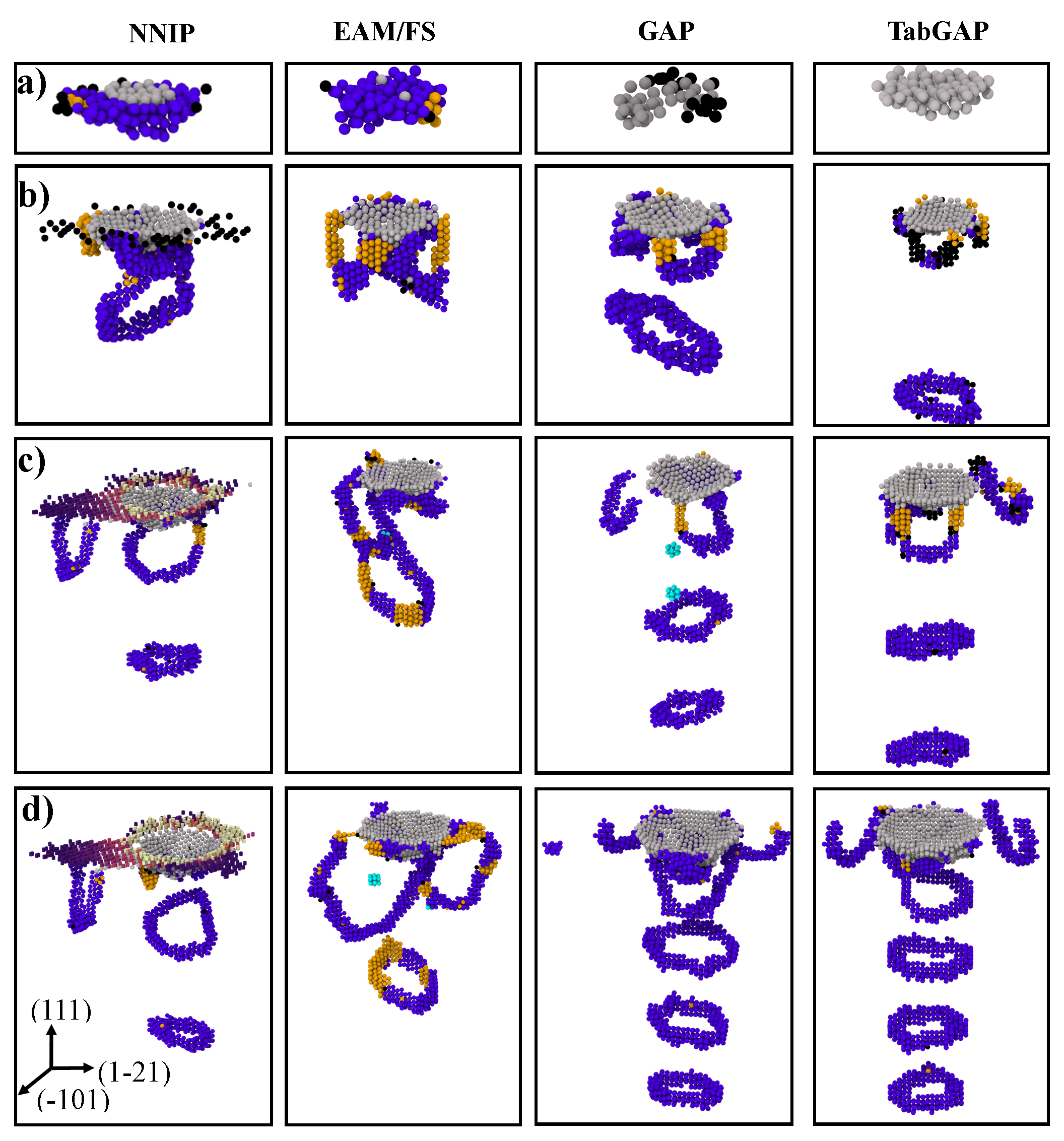}
    \caption{
    Identified defects of indented 
    (111) Mo sample by BDA method at different depths by NNIP, 
    EAM, TabGAP, and GAP approaches. 
    The various defect types are 
    depicted using different colors: gray spheres represent 
    surface atoms in direct contact with the indenter tip, blue 
    spheres indicate edge dislocations, light-blue spheres 
    represent atoms in the vicinity of vacancies, yellow spheres 
    depict twin/screw dislocations, and black spheres highlight 
    unidentified defect atoms.
    The nucleation and propagation of edge dislocations on the 
    \{111\} slip 
    family are observed, which then evolve into prismatic loops.
    In addition, identified slip traces and pile-ups are well 
    modeled by NNIP 
    simulations showing the well--known three-fold symmetric 
    rosette depths below 1.45 nm that are 
    formed by [11$\Bar{2}$], [$\Bar{1}$01] and [0$\Bar{1}$1] 
    planes. }
    \label{fig:bda_method}
\end{figure}

Fig.~\ref{fig:bda_method} illustrates the defects detected 
using the BCC Defect Analysis (BDA)
method (see Methods) in a (111) Mo sample at different depths \cite{moller2016bda}. 
The NNIP nanoindentation simulations in
the initial stages of loading process notably enhance 
the description of the interaction between the indenter tip
and the atoms in the uppermost layers of the surface (see Fig.~\ref{fig:bda_method}(a)). 
In this context, a few Mo atoms located at the very top 
surface layer are recognized as surface defects. 
Additionally, Mo atoms situated beneath these surface defects
begin to coalesce, forming edge dislocations that have 
the potential to evolve into shear loops, contrary to the other simulations where the interatomic 
potentials are not aware of this mechanism.
NNIP simulations are also
anticipated to accurately capture the nucleation and 
propagation of shear loops on \{112\} planes (See Fig.~\ref{fig:bda_method}(b)), as observed experimentally in BCC 
materials \cite{VARILLAS2017431,REMINGTON2014378,MULEWSKA2023154690}. 
Furthermore, NNIP effectively models the nucleation of 
loops through a lasso mechanism, a behavior where GAP and tabGAP 
induced the formation of multiple loops, 
as observed in Fig.~\ref{fig:bda_method}(c)) at a depth of 1.45nm. 
For NNIP, at the maximum indentation depth, it is evident on the sample's surface that displaced atoms align with the slip planes in a 
characteristic three--folded rosette pattern typical for BCC 
materials in the [111] orientation (Fig.~\ref{fig:bda_method}(d)), 
formed by [11$\Bar{2}$], [$\Bar{1}$01], and [0$\Bar{1}$1] 
planes.
In contrast, 
neither GAP nor tabGAP, nor EAM, can adequately incorporate
this description due to the lack of information regarding
pile--up formations.
Besides, the nucleation of more loops is noted, but the 
circumference of the second loop is larger than that of
the first loop. 
The EAM and tabGAP simulations demonstrate a slower and faster 
process, respectively.

In Fig.~\ref{fig:pileups}, we display the atomic
displacement mapping of the [111] Mo sample obtained by
NNIP in (a), EAM/FS in (b), GAP in (c), and tabGAP in
(d) at the maximum indentation depth. The surface of
the sample clearly shows displaced atoms aligned with
the slip planes, forming the characteristic three-fold
rosette pattern typical for BCC materials in the
[111] orientation, as illustrated by the NNIP results
in Fig.~\ref{fig:pileups}(a)). This pattern is created by 
[11$\Bar{2}$], [$\Bar{1}$01], and [0$\Bar{1}$1] planes
\cite{KURPASKA2022110639,VARILLAS2017431}. To assist 
in identifying the shape of the rosette, we have added orange lines, 
reminiscent of what can be observed in SEM images of BCC materials
\cite{REMINGTON2014378}. 
Here NNIP simulations are in good agreement with typical 
observations of pile--up evolution.
However, it is important to note that neither GAP, tabGAP,
nor EAM 
can provide a comprehensive description due to their lack of 
information about open boundary simulation under external loading.

\section{\label{sec:discussion} Discussion 
and conclusions
} 
Interatomic potentials developed before the present work, 
although adequate for many applications, need to be improved for nanoindentation simulations. For example, J. Byggmästar {\em et. al.}~\cite{dataset} developed a GAP potential for Mo, demonstrating accuracy and transferability for elastic, thermal, liquid, defect, and surface properties. However, this potential failed to produce reliable data for the shear stress in the elastic region in the early stages of the nanoindentation simulation. Furthermore, in contrast to the NNIP developed here, 
many prismatic loops were nucleated during the nanoindentation (see Fig.~\ref{fig:bda_method}), potentially due to insufficient information in the energy landscape regarding the dislocation cores, a fact that was illustrated based on the similarity of the GSF configurations with the dislocation cores as depicted in Fig.~\ref{fig:GSFE}. 

The tabGAP potentials are designed for complex multi-element materials \cite{PhysRevMaterials.6.083801}, employing simple low-dimensional descriptors. Although tabGAP potentials have notable accuracy for entropy alloys \cite{PhysRevB.104.104101}, the same issues as the GAP potential arise when it comes to single element BCC Mo. As mentioned earlier, the tabGAP potential leads to nucleation of too 
many prismatic dislocation loops in the nanoindentation simulations (see Fig.~\ref{fig:bda_method}).
Moreover, accurate predictions of shear stress in the initial phases of the nanoindentation simulations were not achieved. 

The EAM/FS potential utilized in the present work ~\cite{ESalonen_2003}, originally designed for radiation damage simulations, failed to accurately produce GSFE curve for Mo in both pristine crystalline and GSF configurations representing dislocation cores. Consequently, it is 
unclear whether or not 
this potential can reliably predict dislocation nucleation during indentation. In addition, similar to the other potentials, it does not correctly predict the nanoindentation shear stress.

Considering the challenges faced in nanoindentation simulations, the presence of a well-developed methodology to tackle these issues would be highly beneficial. 
In this work, we met this goal by introducing to the training dataset
new configurations which  resemble the local atomic environments of an indented sample. 
The similarity measurements presented here ensure the relevance of the newly introduced structures to a nanoindentation simulation. 
To the best of our knowledge, this study represents the first attempt to develop a MLFF specifically 
designed for nanoindentation simulations.  
The novel configurations 
introduced here
could aid the development of
MLFFs for other materials. 

\textbf{\begin{figure}[t!]
    \centering
    \includegraphics[width=0.48\textwidth]{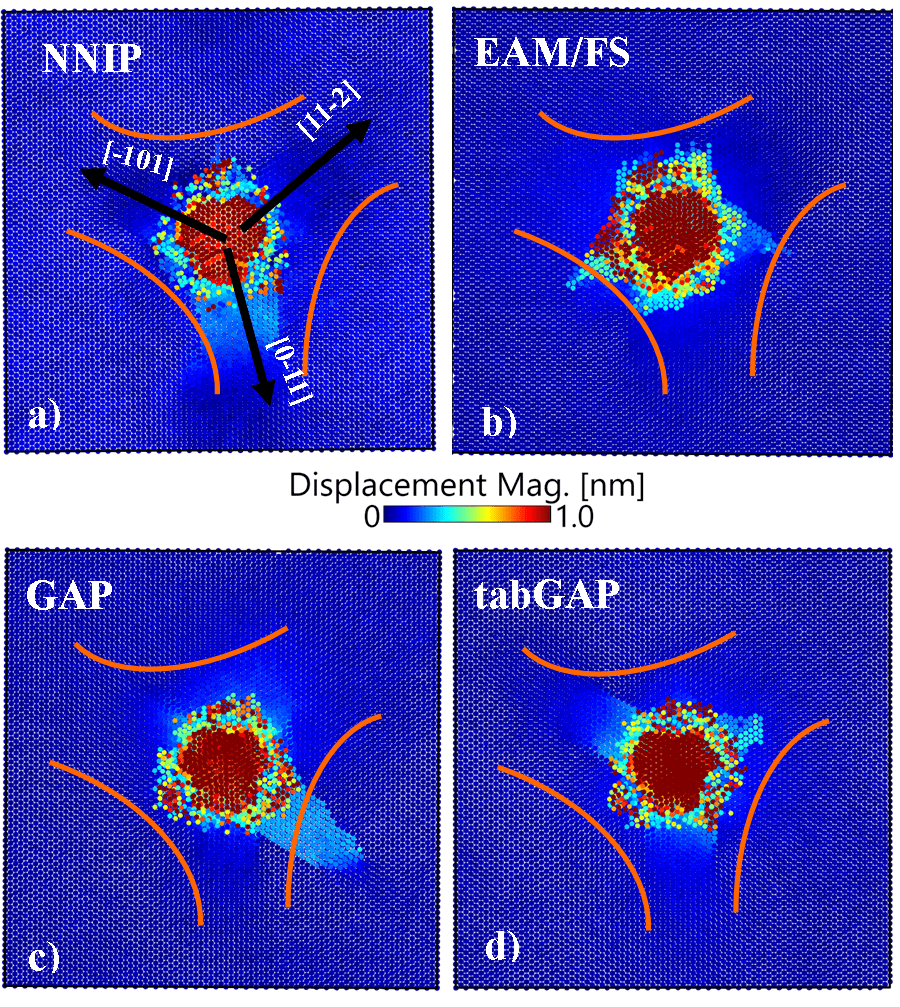}
    \caption{(Color online) Pileups and slip trace formations
    were observed on the $\{112\}$ and $\{011\}$ planes, as 
    well as symmetric ones, for the [111] Mo samples using 
    different methods: NNIP in (a), EAM/FS in (b), GAP in (c),
    and tabGAP in (d). In this analysis, we have included an 
    orange line to emphasize the 3--fold rosette characteristic
    commonly seen in the indentation of BCC samples where NNIP simulations are capable to model it \cite{PhysRevMaterials.7.043603,VARILLAS2017431,DOMINGUEZGUTIERREZ2021141912}. }
    \label{fig:pileups}
\end{figure}}

\section{\label{sec:Methods and Material}Methods}

\subsection{Descriptor parameters}

In this work, PANNA: Properties from Artificial Neural 
Network Architectures~\cite{panna}, which utilizes Tensorflow~\cite{tensorflow2015-whitepaper} to train/evaluate fully-connected feed-forward NNIPs, is used to develop the interatomic potential, with 
the modified version of Behler–Parrinello (mBP) descriptors~\cite{C6SC05720A, PhysRevLett.98.146401}.  
The mBP representation generates a fixed-size vector (the G-vector) for each atom in each configuration of the dataset. Each G-vector describes the environment of the corresponding atom of the configuration to which it belongs, up to a cutoff radius $R_c$. Although higher dimensional G-vectors lead to a more accurate representation of the target potential energy surface, oversized ones increase the MD simulation computational cost. In terms of the distances $R_{ij}$ and $R_{ik}$ of the atom $i$ from  its neighbors $j$ and $k$ and the angle subtended by those distances $\theta_{ijk}$, the radial and angular G-vectors are given by:

\begin{equation}
        G^{rad}_{i}[s]= \sum_{i\neq j} e^{-\eta (R_{ij} - R_s)^2}f_c(R_{ij})
\label{eq:G-1}
\end{equation}
\begin{equation}
\begin{split}
G^{ang}_{i}[s]&=2^{1-\zeta} \sum_{j,k\neq i} [1+cos(\theta_{ijk} - \theta_s)]^\zeta
        \\
    &\times e^{-\eta[\frac{1}{2}(R_{ij} + R_{ik})-R_s]^2}
    f_c(R_{ij}) f_c(R_{ik})
\end{split}
\label{eq:G-2}
\end{equation}
where the smooth cutoff function (which includes the cutoff radius $R_c$) is given by:
\begin{equation}
f_c(R_{ij})= 
    \begin{cases}
    \frac{1}{2}\left [\cos\left (\frac{\pi R_{ij}}{R_c}\right )+1\right ], & R_{ij} \leq R_c\\
    0,   &   R_{ij} > R_c
    \end{cases}
\label{eq:G-3}
\end{equation}
and \(\eta\), \(\zeta\), \(\theta_s\) and $R_s$ are parameters, different for the radial and angular parts. Table.~\ref{tab:T2} shows all values selected for the descriptor parameters in this study.
The choice of the cutoff value is made so that it covers up to three nearest neighbours of the center atom in the BCC Mo, which has a lattice constant of $a = 3.17$ \AA, and thus the third nearest neighbour's distance is $a\times\sqrt{2} = 4.48$ \AA. The length of the G-vector for a single element system is 
\begin{equation} |G_{i}[s]| = (R_s^{ang} \times \theta_s) + R_s^{rad}, 
\end{equation}
which leads to a G-vector of length 152, given the parameters reported in Table.~\ref{tab:T2}. 

\begin{table}[bh!]
\centering
\begin{tabular}{llc}
\hline
Descriptor parameter & Symbol & Value \\
\hline
\textbf{Radial component:}      &                 & \\
Radial exponent (\AA$^{-2}$)       & $\eta^{rad}$ & $32$\\
cutoff (\AA)                  & $R_c^{rad}$  & $5$\\
Number of $R_s$ radial          & $R_s^{rad}$  & $24$\\
\textbf{Angular component:}     &                 &\\
Radial exponent (\AA$^{-2}$)       &$\eta^{ang}$ & $16$\\
cutoff (\AA)                  &$R_c^{ang}$  & $5$\\
Number of $R_s$ angular         &$R_s^{ang}$  & $8$\\
Angular exponent                &$\zeta$          & $128$ \\
Number of $\theta_s$            &$\theta_s$       & $16$\\
\hline
\end{tabular}
\caption{Values of the parameters that appear in the definition of the radial and angular G-vectors, Eq.s\eqref{eq:G-1}, 
\eqref{eq:G-2}, \eqref{eq:G-3}.
Where a number of components is given, the values are equispaced.}
\label{tab:T2}
\end{table}

\subsection{Similarity measurements}

In this study, a distance-based criterion is utilized to quantify the similarity between two distinct configurations. This criterion is subsequently extended to evaluate the closeness of two disparate datasets to one another. Consider two configurations, labeled as $\alpha$ and $\beta$, with $n$ and $m$ atoms per supercell, respectively. The distance matrix for the two configurations, \textbf{$D^{\alpha\beta}$}, has a $n \times m$ dimension and each element of the matrix is the eucleadian distance of atom $i$ in $\alpha$ to atom $j$ in $\beta$:

\begin{equation}
 D^{\alpha\beta}_{i,j} = \norm{G^{\alpha}_{i}[s] - G^{\beta}_{j}[s]}_2
\label{Distance}
\end{equation}

Where $i$ $\in$ $\displaystyle \{1, 2, \dots, n \}$ and $j$ $\in$ $\displaystyle \{1, 2, \dots, m \}$, and each $G[s]$ is a 152 dimensional vector, as explained in the previous section. 
Given this matrix, we can compute the minimum distance of each atom $i$ in configuration $\alpha$ from any atom $j$ in configuration $\beta$, and we define the similarity measure from $\alpha$ to $\beta$ as the maximum among these minima, i.e.:
\begin{equation}
    D^{\alpha\beta}=\max_i\min_j D^{\alpha\beta}_{i,j}.
\end{equation}

It must be noted that this final quantity is not a proper distance, but a non-symmetric quantity giving us 
the ``similarity measure'' method, explained in this section, is subsequently employed to gain insights from the initial dataset. This also includes exploring ways to enhance the dataset through innovative configurations, specifically in relation to an indented supercell. For instance, one can compute the average of all $D^{\alpha\beta}$ values between atoms from two distinct datasets or configuration types within a dataset. This calculation provides an indication of the degree of (dis)similarity between considered datasets/configuration types. The same goes for measuring the similarity of a dataset to a targeted simulation, which in our case is an indented sample.

\subsection{Dataset evaluation and improvement}

As a starting point, we used a dataset~\cite{dataset} originally developed to train a MLFF within the GAP framework~\cite{PhysRevLett.104.136403,PhysRevMaterials.4.093802}. The objective was to 
determine whether this dataset accurately represents the atomic configurations 
occurring during nanoindentation simulations, for which we
employed an EAM potential~\cite{ESalonen_2003}. 
We then analyzed the 
obtained data to determine the degree of similarity between the atomic 
configurations in the dataset and those observed during the nanoindentation 
simulations.

\begin{table}[th]
\centering
\resizebox{0.75\columnwidth}{!}{%
\begin{tabular}{crrr}
\hline
Structure type & $N_s$	& \; \; $N_{at}$    & \; \; $N_{sel}$\\
\hline
Isolated atom		  & 1     & 1      & None\\
Dimer           	  & 19    & 2      & None\\
Slice sample		  & 1996  & 1      & All\\
Distorted BCC         & 547   & 2      & All\\
A15                   & 100   & 8      & None\\
C15                   & 100   & 6      & None\\
HCP                   & 100   & 2      & All\\
FCC                   & 100   & 1      & None\\
Diamond               & 100   & 2      & None\\
Phonon                & 50    & 54     & All\\
Self-interstitials (SIA)    & 32    & 121    & 14\\
di-Self-interstitials & 14    & 122-252& All\\
Simple Cubic          & 100   & 1      & None\\
Vacancy               & 210   & 53     & All\\
di-Vacancy            & 10    & 118    & All\\
tri-Vacancy           & 14    & 117    & All\\
Liquid                & 45    & 128    & None\\
Short range           & 90    & 53-55  & None\\
Surface (100)         & 45    & 12     & All\\
Surface (110)         & 45    & 12     & All\\
Surface (111)         & 41    & 12     & All\\
Surface (112)         & 45    & 12     & All\\
Liquid Surface        & 24    & 128    & All\\
$\gamma$-urface         & 178   & 12     & All\\
\textbf{GSFCs}        & 100   & 18     & All \\
\textbf{GSFCs + SIA}  & 100   & 55     & All \\
\textbf{Pileup}       & 1000  & 32     & All \\
\textbf{HT + substrate} & 600 & 54-72  & All \\

\hline
\end{tabular}
}
\caption{The original dataset from~\cite{dataset}, showing $N_s$ as the number of structures, $N_{at}$ as the number of atoms per configuration, and $N_{sel}$ as the number of selected configurations in the final dataset.}
\label{tab:configs}
\end{table}

\begin{figure*}[ht]
  \centering
  \includegraphics[width=0.98\textwidth]{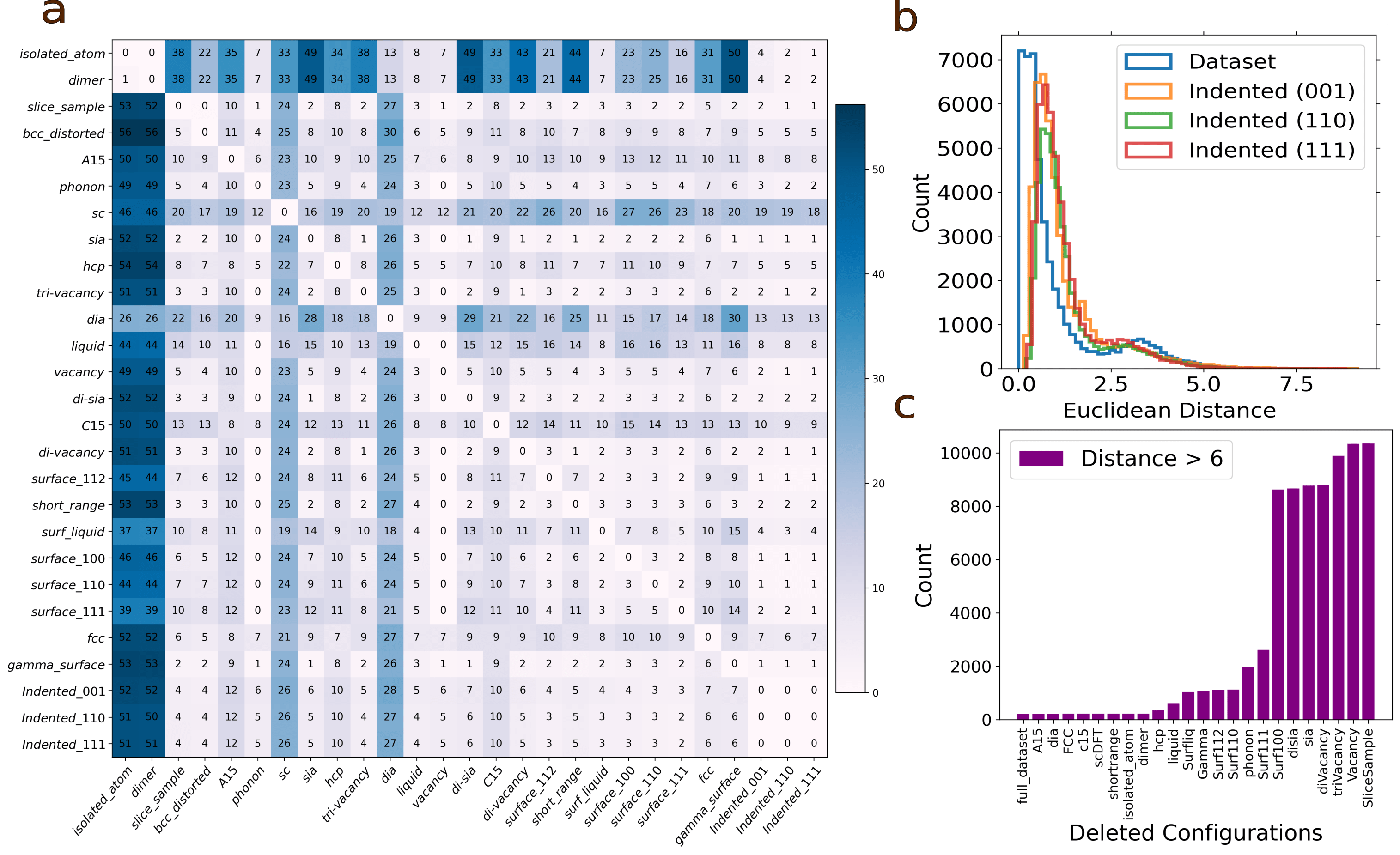}
  \caption{Dataset evaluation and stability. a) Average of minimum distances between different configuration types as well as with the indented samples. b) Distribution of the minimum distances of different configurations to the rest of the data points in the dataset and the minimum distances of the dataset with the indented configurations. c) Effect of configuration removal from the dataset on the minimum distances of the dataset to the indented sample.} 
  \label{fig:originaldataset}
\end{figure*} 

This level of similarity is evaluated by identifying which atom $j$ in the dataset has the minimum distance to each atom $i$ in the indented sample. The obtained value corresponds to the \emph{largest minimum} distance for each atom in the indented sample from all the atoms in the dataset. The concept of ``distance'' for two atoms $i$ and $j$, refers to the $l^{2}$-norm of $G_{i}[s] - G_{j}[s]$, where $G[s]$ are the fixed-size modified Behler-Parinnelo (mBP) descriptor vectors~\cite{C6SC05720A, PhysRevLett.98.146401}, as discussed in the previous sections.
To this end, we compared the configuration 
types present in the dataset to those of all atoms identified in the indented 
sample and drew conclusions based on the level of correspondence between the two sets. 
Through this analysis, we aimed to gain insights into the suitability of the 
selected dataset for studying nanoindentation behavior and identifying the 
underlying mechanisms governing it.

To generate a suitable dataset for training a NNIP targeted at nanoindentation simulations, it is crucial to ensure that the configurations included accurately represent the three essential regions of a sample under indentation. These regions include the atoms on the surface of the sample, which correspond to the pileup patterns, the atoms situated beneath the indenter tip that undergo significant plastic deformation, and the atoms located on the nucleated dislocation cores. The 
evaluation of these three critical regions and 
development of configurations that closely resemble them can 
serve as a 
benchmark
for ML potentials for other BCC materials.

Before comparing the original dataset with the indented sample, we calculated the average minimum distance between each pair of configuration types and generated a correlation figure to visualize their proximity (Fig.~\ref{fig:originaldataset}(a)).  
It is evident from this figure that although the isolated atom and dimer configurations are quite distant from almost all other configurations, they are relatively close to the indented samples. However, these configurations were not included in the final dataset due to their low numbers (1 and 19, respectively), which were deemed insufficient for training a NNIP. Moreover, the A15, simple cubic (sc), diamond (dia), liquid and C15 configurations were removed from the final dataset as they were located at a distance beyond the set threshold from the indented configurations, with simple cubic, diamond, and C15 configurations having the largest distance. Furthermore, we excluded short-range configurations from the final dataset because their energies varied significantly (Fig.~\ref{fig:s1}(a)), leading to training difficulties. Finally, to reduce computational cost, we kept only half of the self-interstitial configurations in the final dataset. Table.~\ref{tab:configs} summarizes all modifications made to the original dataset.

Several methods can be employed to determine a ``good" threshold for deciding whether to keep or remove a particular configuration from the dataset, based on its similarity to the indented configuration. In this study, we have chosen to use the start of the tail of the distribution of the minimum values in the dataset distance matrix as the threshold, which is approximately 6 based on Fig.~\ref{fig:originaldataset}(b). Fig.~\ref{fig:originaldataset}(b) also demonstrates that this value is consistent with the minimum distances between the dataset configurations and all three orientations of the indented samples. All decisions regarding whether to keep or remove a configuration from the final dataset in this study are based on this threshold.

To ensure the accuracy of the modifications made to the dataset, we removed one type of configuration from the dataset at a time and quantified the number of atoms in the indented samples that had minimum distances greater than 5 from the dataset (Fig.~\ref{fig:originaldataset}(c)). As our analysis shows, the number of atoms with a minimum distance greater than 5 to the dataset does not increase when A15, diamond, Face-Centered Cubic (FCC), simple cubic, short range, isolated atom, and dimer configurations are removed, indicating the dataset's stability against the indented samples, whether these configurations are present in the dataset or not. However, upon removing Hexagonal Close-Packed (HCP) configurations, the number of atoms with a large distance from the dataset increases, which is consistent with the fact that the average minimum distance of HCP configurations to the indented samples is 5 . The greatest increase in the number of atoms with a distance greater than 5 ~ from the dataset occurs when surface configurations are removed, which underscores their importance since they represent the surface in the nanoindentation simulations.

Following the modifications made to the dataset obtained from \cite{dataset}, attempts were made to enhance its quality by incorporating various types of configurations and reevaluating the distances of the indented configuration from the dataset. Among the crucial local environments that should be included in the dataset are the atoms belonging to the dislocation cores. Furthermore, it was discovered that the distances of the atoms beneath the indenter tip and those on the surface exceeded the selected 6 threshold (Fig.~\ref{fig:newconfigs}(a)). These environments in the indented samples are critical to be covered in the dataset since dislocation cores play a vital role in the dislocation dynamics properties, and the atoms beneath the indenter tip trigger these line defects. Additionally, the plastic region beneath the surface is responsible for the pile-up patterns that appear on the surface of the indented configuration. It is also imperative to incorporate configurations in the dataset representing atoms on the surface to capture this phenomenon.

In order to model the aforementioned regions of an indented sample, we explored the use of high-temperature configurations to effectively reduce the distance between the atoms in these areas and the dataset, as depicted in Fig.~\ref{fig:newconfigs}(b). Nevertheless, accounting for the atoms beneath the indenter tip and on the surface requires the inclusion of a layer of frozen atoms in the high-temperature configurations that emulate the contact of the sample with the indenter tip. The addition of $1000$ isothermal-isobaric ensemble (NPT) high-temperature configurations with $16$ atoms in the $2 \times 2 \times 2$ supercells appeared to decrease the distances between the atoms on the dislocation cores and the dataset. We utilized the same approach for the atoms beneath the indenter tip. In this regard, we introduced $600$ configurations, denoted as ``high temperature $+$ substrate", where a layer of atoms was frozen while other atoms were heated to high temperatures (under the melting point). While there are various unique layers of atoms that can be taken into account as the contact to the substrate, we verified that 300 configurations ($3 \times 3 \times 3$) with $54$ atoms per supercell---displayed in the middle figure of Fig.~\ref{fig:newconfigs}-(b)---were adequate and most relevant after trying different layers of atoms. Additionally, we included 300 configurations ($4 \times 3 \times 3$) with $72$ atoms per supercell, which featured a layer of atoms frozen on top. Moreover, it is crucial for a NNIP's dataset to incorporate configurations resembling the atoms located on the surface of the indented sample. To achieve this, we introduced $1000$ BCC surface configurations ($3 \times 3 \times 2$) with $32$ atoms per supercell, where a layer of atoms was frozen on top while the remaining atoms were subjected to high temperature. These configurations were named ``pileup" in our study and enabled the dataset to account for the atoms in this region.

Another effective approach to cover the dislocation cores is through the use of GSFCs that incorporate a self-interstitial atom (SIA) atom on the surface (as illustrated in Fig.~\ref{fig:GSFCs}(a)). These configurations have been found to be particularly effective in reducing the distances of atoms on the dislocation cores from the dataset. While the use of GSFCs without an SIA on the surface can also reduce distances of atoms beneath the dislocation cores and on the slipping plane (as shown in Fig.~\ref{fig:GSFCs}(b)), it may not entirely cover all the atoms on the dislocation core.

Incorporating a SIA on the surface of the GSFCs leads to a significant decrease in the distances of almost all atoms on the dislocation cores from the dataset (as demonstrated in Fig.~\ref{fig:GSFCs}(c)). Notably, the use of GSFCs with a SIA instead of high-temperature configurations solely for the dislocation cores presents several advantages. For instance, the distribution of energies of these configurations is narrower, facilitating the learning process for the network (as depicted in Fig.~\ref{fig:s1}(b)). Furthermore, only $100$ GSFCs with SIA configurations, as opposed to the $1000$ mentioned for high-temperature configurations, can ease the process of training
the network. Additionally, using GSFCs with SIA configurations guarantees that no atoms on the dislocation core will have a distance greater than 6, thus ensuring the closeness of the distances of these configurations to the dislocation cores.

The visualization of the distances between the atoms in all three regions of interest and the dataset reveals a significant reduction in distances after incorporating the high-temperature configurations (see Fig.~\ref{fig:newconfigs}(c)). The distribution of distances for each region before and after adding the high-temperature configurations is depicted in Fig.~\ref{fig:newconfigs}(d). 
Although a few atoms still have distances greater than  6 under the indenter tip, the number of such atoms has notably decreased after adding the appropriate configurations. Finally, because we are trying to develop  a NNIP for the case of nanoindentation simulation during which atoms are compressed under the indenter tip, we added $300$ compressed $3 \times 3 \times 4$ configurations with each of them including 72 atoms.

\subsection{DFT calculations}

The DFT calculations were performed with the \texttt{Quantum Espresso} \cite{Giannozzi_2009, Giannozzi_2017} (QE) package, using a norm-conserving PBEsol exchange-correlation functional \cite{Prandini2018, doi:10.1126/science.aad3000, PhysRevB.88.085117} and 14 valence electrons. The Brillouin zone was sampled using Monkhorst-Pack method \cite{PhysRevB.13.5188}, and, from the convergence analysis of Fig.~\ref{fig:s2}, the k-point mesh and plane-wave cutoff energy in a Mo unit-cell were set to $8\times8\times8$ and 60 $Ry$, respectively. The selected k-point grid was rescaled for supercells calculations according to their dimension, implying the use of a $2\times2\times2$ grid for $4\times4\times4$ conventional super-cells, and was set to $1\times1\times1$ for any bigger configuration. Smearing was introduced within the Methfessel-Paxton method \cite{PhysRevB.40.3616} to help convergence, with a spreading of 0.00735 $Ry$ (0.1 $eV$). The structural properties, involving elastic constants $C_{ij}$, Bulk modulus $B$ (in the Voigt-Reuss-Hill approximation~\cite{chung1967voigt}) and Poisson ratio $\nu$, have been computed running the QE driver \texttt{THERMO\_PW}~\cite{thermopw} on a Mo unit-cell.

The total energies of the configurations obtained from~\cite{dataset} were compared with the values calculated in our work to make sure of their consistency, which is shown in Fig.~\ref{fig:s4}. 

\subsection{Neural Network Training}

In the PANNA framework, the environmental descriptors of each atom are provided as input to a fully connected network with two hidden layers, consisting of 256 and 128 nodes for the first and second layers, respectively, both with Gaussian activation function, and a single-node output layer with linear activation. The atomic environment is represented by a descriptor with 152 components, resulting in a network with 71808 weights and 385 biases.
A batch size of $10$ 
is utilized for training, while the model is trained using initial random weights and a constant learning rate of $10^{-4}$ throughout the training process. In this methodology, the energy of a configuration consisting of $N$ atoms is defined as the sum of atomic energy contributions:
\begin{equation}\label{energy}
E = \sum_{i=1}^NE_i(G_i),
\end{equation}
where $E_i$ is the energy of atom $i$ with a G-vector of $G_i$. 
The force on atom $i$ 
which is situated 
at position $\vec{R}_i$
is given by:
\begin{equation}\label{force}
\vec{F}_i = -\sum_{j}\sum_{\mu}\frac{\partial E_j}{\partial G_{j\mu}}\frac{\partial G_{j\mu}}{\partial \vec{R}_i}
\end{equation}
with $j$ labeling the atoms located within the cutoff distance of atom $i$
and $\mu$ labeling the descriptor components. 

To optimize the weights and bias parameters of the network, we use the Adam algorithm~\cite{kingma2017adam} 
to compute gradients of randomly selected batches of the training dataset. 
The loss function for optimizing the network weights, 
denoted collectively as $W$, consists of 
two terms, one for 
the energy $\mathcal{L}_E(W)$, 
and one for the forces, 
$\mathcal{L}_F(W)$:
\begin{equation}\label{losstotal}
    \mathcal{L}(W) = \mathcal{L}_E(W) + 
    \mathcal{L}_F(W).
\end{equation}
The energy contribution is given by:
\begin{equation}\label{lossenergy}
    \mathcal{L}_E(W) = \sum_{s \in {\rm batch}} \left[ E_{s}^{{\rm DFT}} - E_{s}(W) \right]^2 
\end{equation}
where $s$ refers to 
the atomic configuration, 
$E_{s}^{DFT}$ 
is the total energy calculated from DFT (the target value) 
and $E_s(W)$ is the total energy predicted by the NNIP. 
The force contribution is given by:
\begin{equation}\label{lossforce}
    \mathcal{L}_F(W) = \lambda_F \sum_{s \in {\rm batch}} \sum_{i=1}^{N_s} \left\lvert \vec{F}_{i;s}^{{\rm DFT}} - \vec{F}_{i;s}(W) \right\rvert^2
\end{equation}
with $\vec{F}_{i;s}^{{\rm DFT}}$ the force obtained from DFT 
and $\vec{F}_{i;s}$ 
the force obtained from the NNIP, 
for atom $i$ in configuration $s$; 
$N_s$ is the total number of atoms in configuration $s$. 
The parameter $\lambda_F$  
adjusts the 
relative contribution of the force component 
and was set to 
$\lambda_F =0.5$. 

\subsection{Nanoindetation simulations}

\subsubsection{Simulation method and parameters}
To establish boundary conditions along the depth ($dz$) of the Mo samples, we 
divided them into three sections in the $z$ direction during the initial stage: 
a frozen section with a width of approximately 
0.02$\times dz$, which ensured numerical cell stability; a thermostatic section 
about 0.08$\times dz$ above the frozen section, which dissipated heat generated 
during nanoindentation; and a dynamical atoms section, where the interaction with 
the indenter tip modified the surface structure of the samples. 
Furthermore, we 
included a 5 nm vacuum section at the top of the sample as an open boundary 
\cite{KURPASKA2022110639}.
We considered the indenter tip as a non-atomic repulsive imaginary (RI) rigid 
sphere and defined its force potential as 
\begin{equation} 
F(t) = K \left(\vec r(t) - R 
\right)^2,
\end{equation} 
where $K = 236$ eV/\AA$^3$ (37.8 GPa) was the force constant, and 
$\vec r(t)$ was the position of the center of the tip as a function of time, with 
a radius $R = 3$ nm. We conducted molecular dynamics (MD) simulations using an 
NVE statistical thermodynamic ensemble and the velocity Verlet algorithm to 
emulate an experimental nanoindentation test. The $x$ and $y$ axes had periodic 
boundary conditions to simulate an infinite surface, while the $z$ orientation 
had a fixed bottom boundary and a free top boundary in all MD simulations 
\cite{DOMINGUEZGUTIERREZ2021141912, PhysRevMaterials.7.043603}.

In our simulations, we chose $\vec r(t) = x_0 \hat x + y_0 \hat y + 
(z_0 \pm vt)\hat z$, where $x_0$ and $y_0$ were the center of the surface 
sample on the $xy$ plane, and $z_0 = 0.5$ nm was the initial gap between 
the surface and the indenter tip. The tip moved with a speed of 
$v = 20$ m/s with a 
time step of $\Delta t = 1$ fs. We chose the maximum indentation depth 
to be 2.0 nm to avoid the influence of boundary layers in the dynamical
atoms region.

\subsubsection{Normalized maximum shear stress}

The contact pressure, $P$, is calculated by using a linear elastic
contact mechanics formulation 
\cite{PhysRevMaterials.7.043603,VARILLAS2017431}:
\begin{equation}
    P = 2\pi  
    \left[ 24p \left( \frac{E_{\rm Y}R}{1-\nu^2} \right)^2 \right]^{1/3},
    \label{eq:pressure}
\end{equation}
with $E_Y$ as the Young's modulus, $p$ as the simulation load, $R$
the indenter radius, and $\nu$ the Poisson's ratio;
the radius of the contact area is obtained with the geometrical 
relationship: 
\begin{equation}
\mathrm{a}(h) = \left[ 3PR \left( 
 1-\nu^2\right)/8E_{\rm Y} \right]^{1/3}
 \end{equation}
 which is related to the  inner radius 
 of the plastic region where the defects nucleate.
 This quantity provides an intrinsic measure of the surface resistance 
 to a specific defect nucleation 
 process~\cite{VARILLAS2017431,PhysRevMaterials.7.043603}.
To determine the strength and stability of the Mo
matrix under load, 
we compute 
the principal stress applied on the $z$ direction
as \cite{REMINGTON2014378} :
\begin{equation}
    \sigma_{zz}  = 
    - \mathcal{S}
    \left[ \left( 1-
\frac{
    \arctan 
    (\alpha)}
    {\alpha}
    \right) 
    (1+\nu)
   -\frac{1}{2(1+
   1/\alpha^2)
   }  \right],   
\end{equation}
where the quantities 
$\mathcal{S}$ and $\alpha$ are defined as:
\begin{equation}
    \mathcal{S} =
    \frac{3P_{\rm ave}}{2 \pi \mathrm{a}(h)^2},
    \quad
    \alpha = 
    \frac{\mathrm{a}(h)}{h}.
    \nonumber
\end{equation}
with $h$ as the indentation depth and $a(h)$ the contact area between 
the indenter tip and the top atomic layers.
The stress applied in the direction parallel to the indenter 
surface is then expressed as:
\begin{equation}
    \sigma_{xx}  =  \sigma_{yy} = - 
\frac{\mathcal{S}}{1+1/\alpha^2
}
\end{equation}
This gives the maximum shear stress: 
\begin{equation}
\tau_{\rm max} =
\frac{1}{2} 
\left( \sigma_{zz}-\sigma_{xx} \right),
\label{eq:shearMax}
\end{equation}
that the material can withstand 
before it begins to undergo plastic deformation, being normalized by the applied pressure (equal to the applied 
force
$F$ divided by the contact area). 
The normalized depth is the distance from the surface of the material to the 
point at which the maximum shear stress occurs, 
normalized by the radius of the indenter that is used 
to apply the shear forces. 

\subsubsection{Defect analysis}

In order to identify the defects in nanoindentation 
simulations, we apply the BCC Defect Analysis (BDA) 
developed by M\"oller and Biztek~\cite{moller2016bda} 
which utilizes coordination number (CN), centrosymmetry 
parameter (CSP), and common neighbor analysis (CNA) 
techniques to detect typical defects found in
bcc crystals. 
The characterization of the materials defects starts 
by calculating CN, CSP, and CNA values of all the atoms by 
considering a cutoff radius of $(1+\sqrt{2})/2a_0$ with $a_0$ 
as the lattice constant of Mo.
Thus, the six next--nearest neighbors of perfect bcc atoms
are into this cutoff and their CN value increases from 8 to 14. 
Consequently, BDA compares the CN and CSP values of each atom 
generating a list of non--bcc neighbors with 
CNA$\neq$bcc and CN$\neq$14 that classifies for the following 
typical defects: surfaces, vacancies, twin boundaries, screw 
dislocations, \{110\} planar faults, and edge dislocations.

\subsubsection*{Author Contributions}
N.D.A. created and designed the DFT training dataset, performed all the NNIP training, compiled and analyzed the data, did NNIP-MD nanoindentation and GSFE simulations and wrote the manuscript. N.D.A. prepared all the figures, jointly with F.J.D. for the nanoindentation section. F.P. and E.K\"u. supervised the training procedure and designed the similarity measurement method. F.J.D. supervised the nanoindentation part. D.M. did the DFT GSFE calculations and DFT validation of elastic properties with \texttt{THERMO\_PW}. S.P. and E.K. supervised all aspects of the work. All authors contributed to revision of the manuscript.

\subsection*{ACKNOWLEDGEMENTS}
We acknowledge support from the European Union Horizon 2020 
Research and 
Innovation Programme under grant agreement no. 857470 and from the European 
Regional Development Fund via the Foundation for Polish Science 
International Research Agenda PLUS program grant No. MAB PLUS/2018/8.  
Computational resources were provided by the High Performance 
Cluster at the National Centre for Nuclear Research in Poland.

\subsubsection*{Competing Interests}
The authors declare no competing interests.




\section{References}

\bibliography{Ref}
\clearpage

\input{sections/SM.tex}

\end{document}

%% file: macros.tex
\definecolor{light-gray}{gray}{0.85}

\newcommand{\drawSlope}[6]{ 
				\coordinate (center1) at (#1,#2); 
				 \FPeval{\nx}{cos(#4*pi/180)}%
				 \FPeval{\ny}{sin(#4*pi/180)}%
				 \FPeval{\absNx}{abs(\nx)}%
				 \FPeval{\absNy}{abs(\ny)}%
				\coordinate (b) at ($ (center1) + #3*(-\ny,+\nx) $);
				\coordinate (c) at ($ (center1) - #3*(-\ny,+\nx) $); 
				 \FPeval{\xx}{(-\nx*\ny)}%
				\ifdim\xx pt < 0pt 
				\coordinate (d) at ($ (c) -2*#3*\ny*(1,0) $);
				\node at ($(d) +(0,#3*\nx)-0.2*(\nx/\absNx,0)$) {{\color{#5}$#6$}}; 
				\node at ($(d)+(#3*\ny,0)-0.2*(0,\ny/\absNy)$) {{\color{#5}$\scriptstyle 1$}}; 
				\else
				\coordinate (d) at ($ (c) +2*#3*\nx*(0,1) $);
				\node at ($(d)-#3*\nx*(0,1)-0.2*\nx/\absNx*(1,0)$) {{\color{#5}$#6$}}; 
				\node at ($(d)-#3*\ny*(1,0)-0.2*\ny/\absNy*(0,1)$) {{\color{#5}$\scriptstyle 1$}}; 
				\fi
				\draw[#5,line width=0.1mm] (d) -- (b); 
				\draw[#5,line width=0.1mm] (d) -- (c); 
				\draw[#5,line width=0.1mm] (b) -- (c); 
} 



%% file: sections/SM.tex
\renewcommand{\thefigure}{S\arabic{figure}}
\renewcommand{\thesection}{S~\Roman{section}}
\setcounter{figure}{0}    
\setcounter{section}{0}    
\setcounter{page}{1}

\begin{titlepage}
  \begin{center}
    \Large\bfseries Supplementary Information for: Neural Network Interatomic Potentials For Open Surface Nanomechanics Applications
  \end{center}

  \begin{center}
    Amirhossein D. Naghdi$^{1,2,*}$, Franco Pellegrini$^{3}$, Emine K\"uc\"ukbenli$^{4,5}$, Dario Massa$^{1,2}$, F. Javier Dominguez--Gutierrez$^{1}$, Efthimios Kaxiras$^{5,6}$, and Stefanos Papanikolaou$^{1,*}$ 
  \end{center}

  \begin{center}
    $^{1}$NOMATEN Centre of Excellence, National Center for Nuclear Research, ul. A. Sołtana 7, 05-400 Swierk/Otwock \\
    $^{2}$IDEAS NCBR, ul. Chmielna 69, 00-801, Warsaw, Poland \\
    $^{3}$International School for Advanced Studies (SISSA), Via Bonomea, 265, I-34136 Trieste, Italy \\
    $^{4}$ Nvidia Corporation, Santa Clara, CA, USA \\
    $^{5}$ John A. Paulson School of Engineering and Applied Sciences, Harvard University, Cambridge, Massachusetts 02138, USA\\
    $^{6}$Department of Physics, Harvard University, Cambridge, Massachusetts 02138, USA\\
  \end{center}

\begin{figure*}[h]
  \centering
  \includegraphics[width=0.80\textwidth]{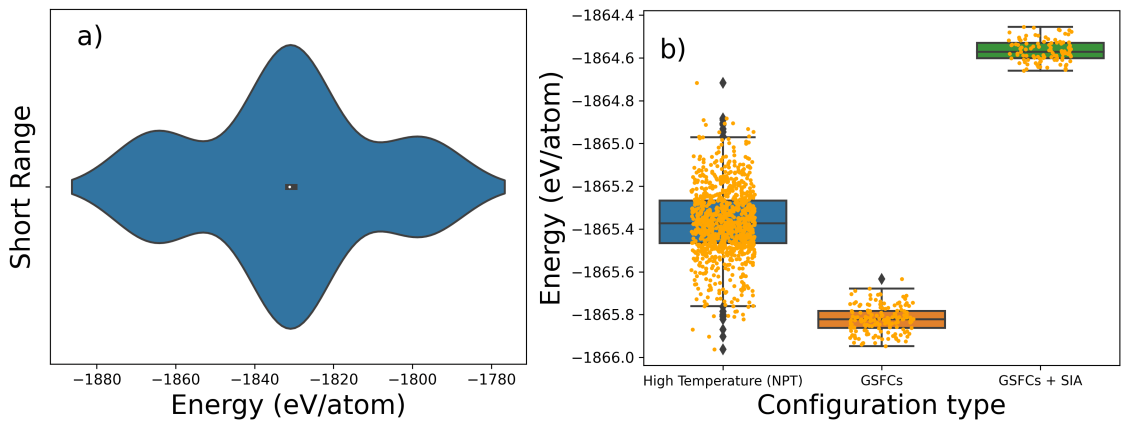}
  \caption{Energy distribution of \textbf{a)} Short range configurations and \textbf{b)} High temperature and GSFCs. The drastic variation of energies for short range configurations (100 meV/atom) suggest that these configurations would be problematic for training a NNIP. n addition, it is shown that although high temperature NPT configurations have a larger energy distribution than
the GSFCs.} 
  \label{fig:s1}
\end{figure*}

\begin{figure*}
    \centering
    \begin{overpic}[width=0.45\textwidth]{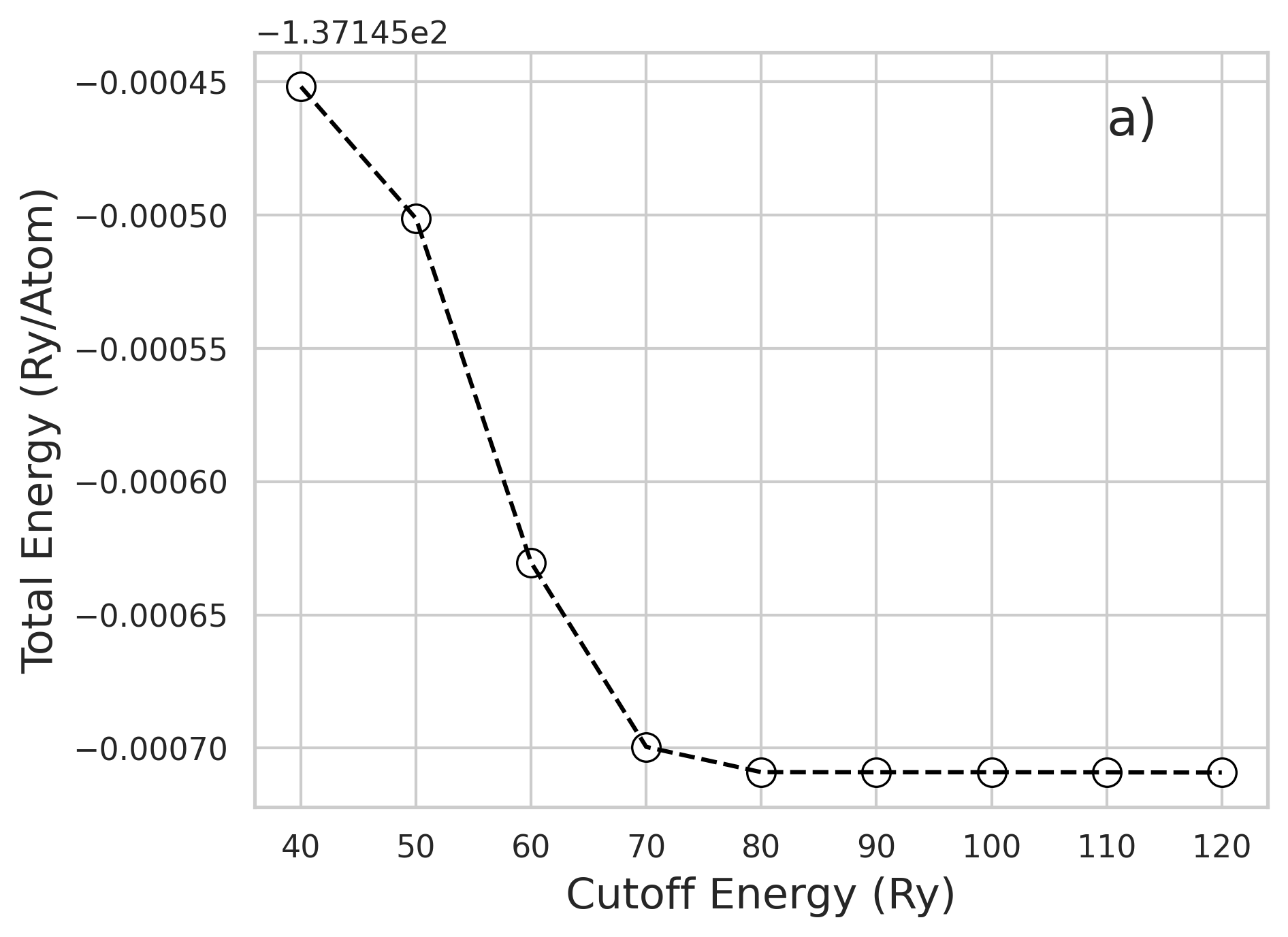}
    \end{overpic}
    \begin{overpic}[width=0.45\textwidth]{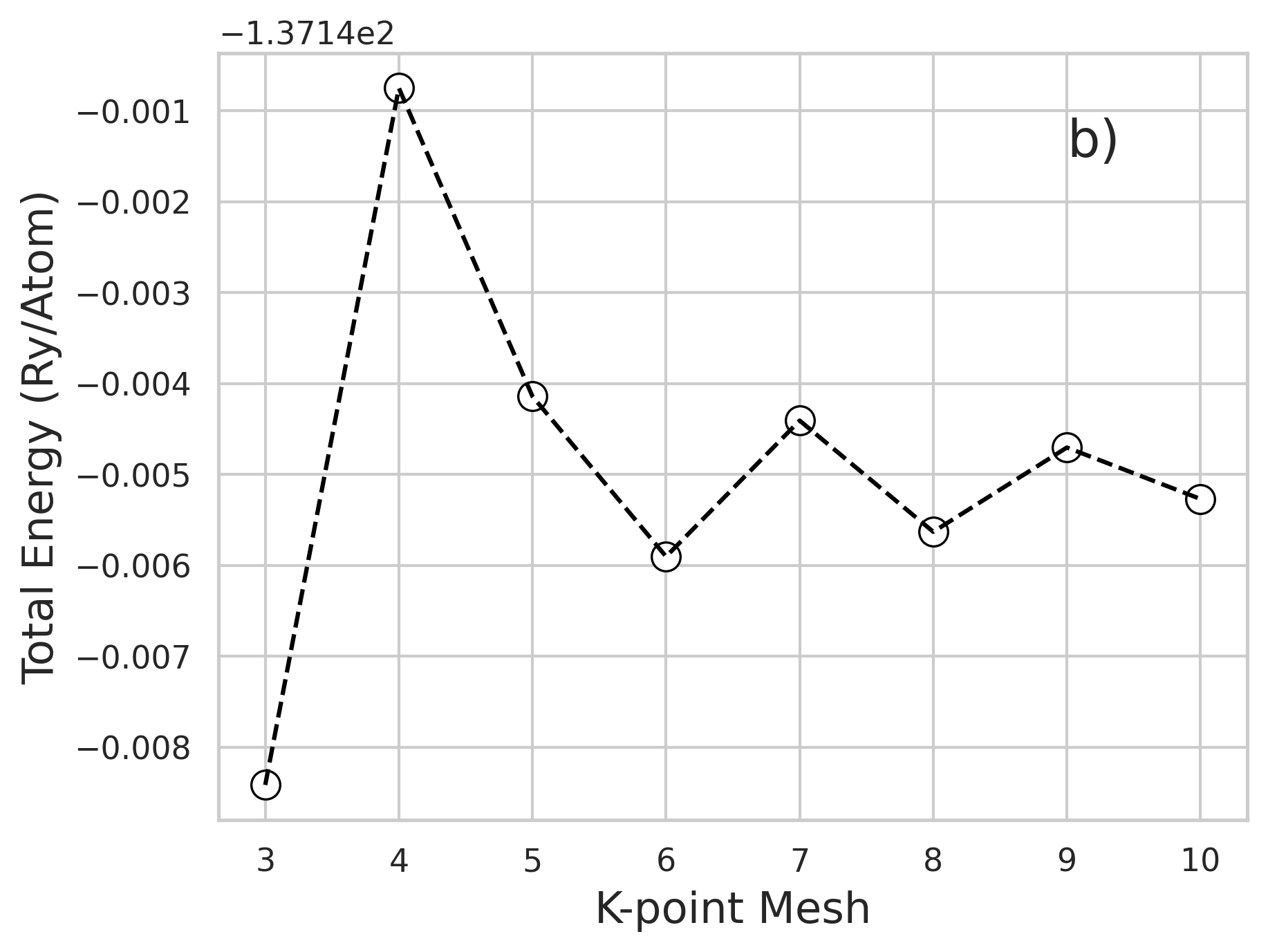}
    \end{overpic}
    \caption{DFT convergence check for \textbf{a)} cutoff energy and \textbf{b} k-point mesh grid.} 
    \label{fig:s2}
\end{figure*}

\end{titlepage}

\def\thefootnote{$*$}
\footnotetext{Corresponding authors\\A.D.N., E-mail: \url{Amirhossein.Naghdi@ncbj.gov.pl}\\S.P., E-mail: \url{Stefanos.Papanikolaou@ncbj.gov.pl}}
\def\thefootnote{\arabic{footnote}}

\begin{figure*}
  \centering
  \includegraphics[width=0.80\textwidth]{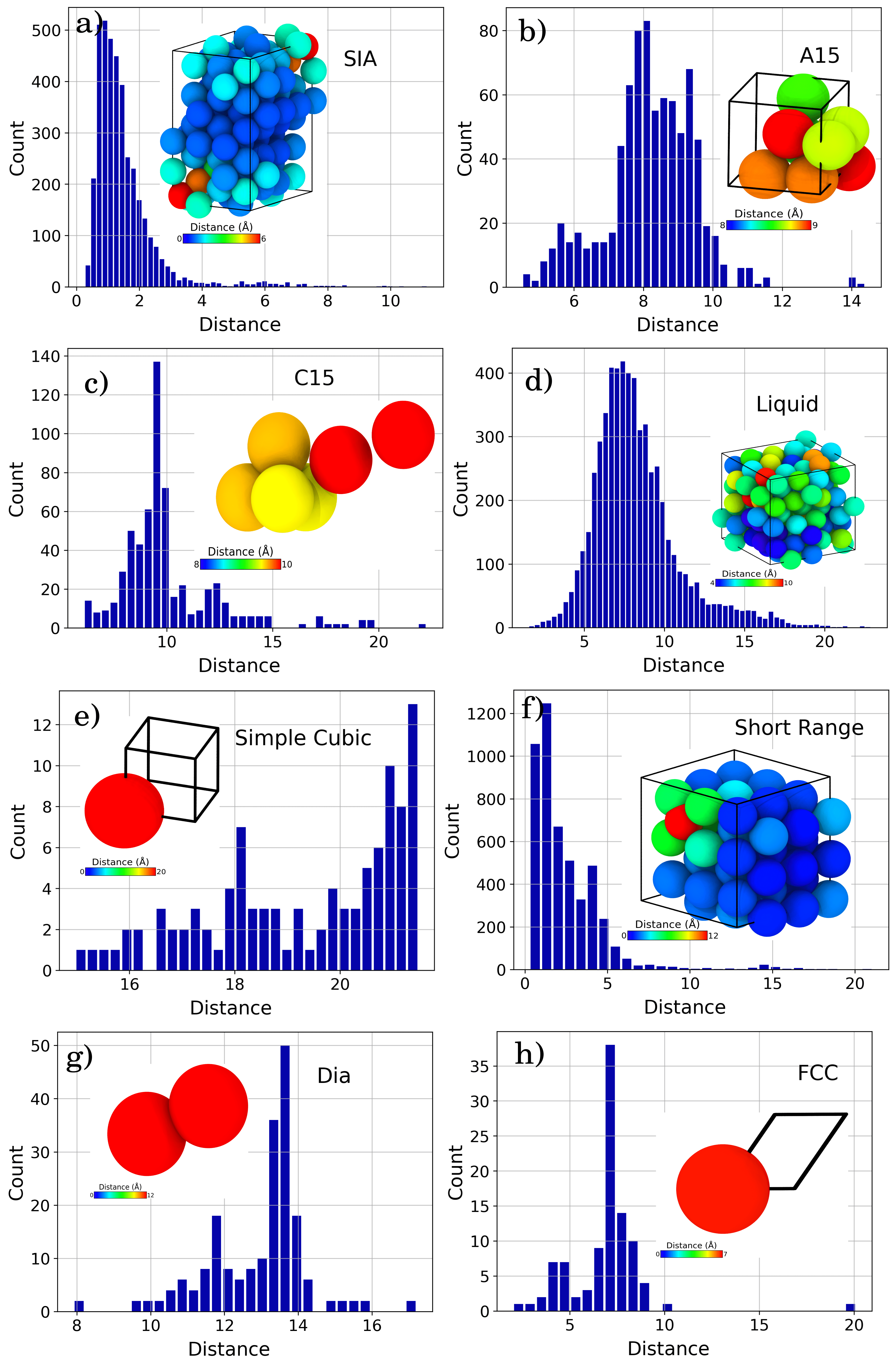}
  \caption{Distribution of minimum distances of each configuration to the indented sample for deleted configurations from the dataset. As it is shown in the insets, each configuration consists of number of atoms with large minimum distance.} 
  \label{fig:s3}
\end{figure*}

\begin{figure*}
  \centering
  \includegraphics[width=0.68\textwidth]{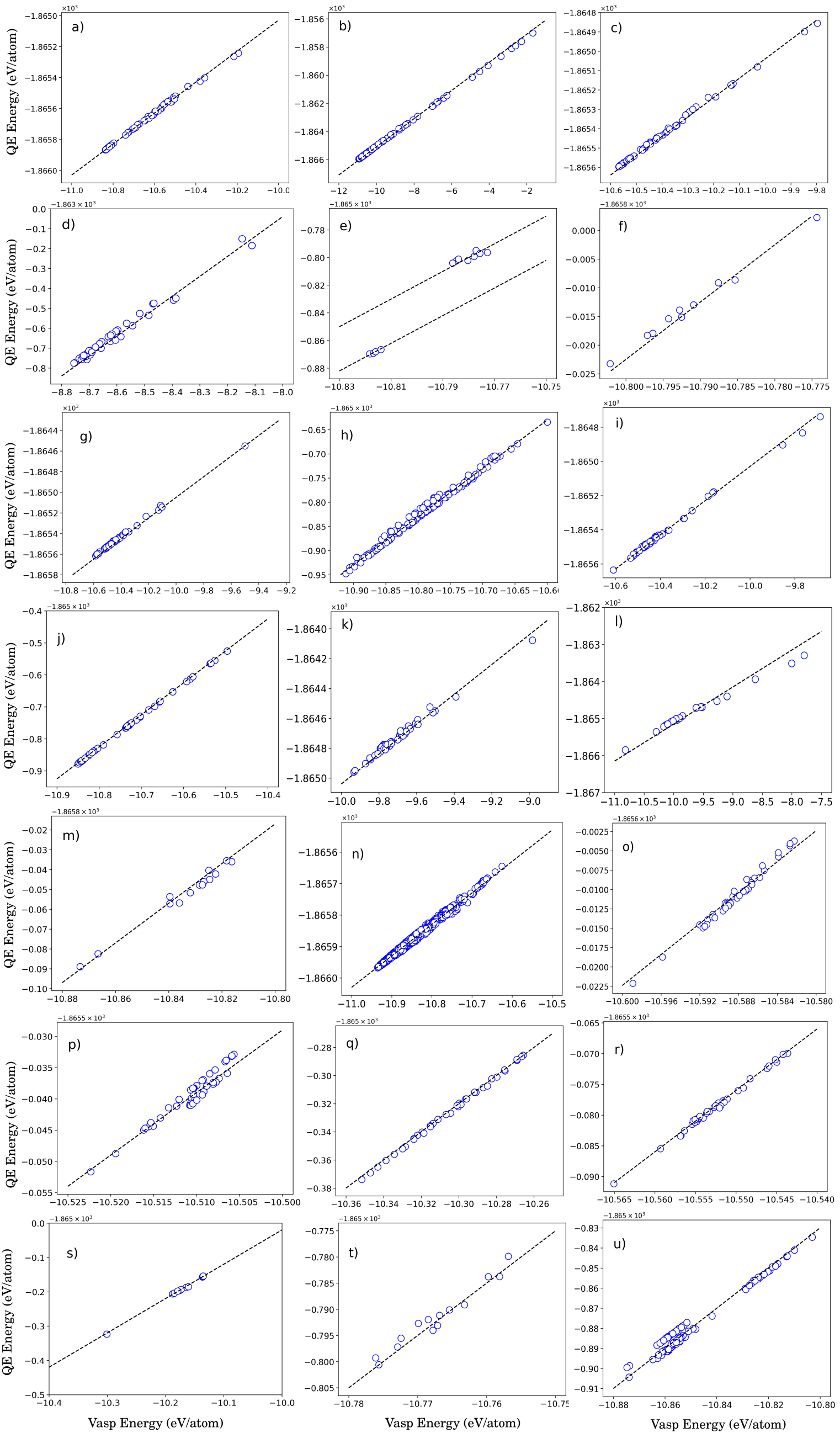}
  \caption{The total energies obtained in this work with Quantum Espresso (QE), compared to VASP (from the original GAP potential). The dashed lines represent a (y = x + b) function with y being the Quantum Espresso energy, x being the VASP energy and b is the y-intercept which is different for each panel and illustrates the correctness of the values we obtained from QE. Figures \textbf{a)} to \textbf{u)} are for the following configurations, respectively: A15, BCC, C15, diamond, di-sia, di-Vacancy, FCC, Gamma, HCP, phonon, sc, short range, sia, sliced sample, surface 100, surface 110, surface 111, surface 112, liquid surface, tri-Vacancy and vacancy.} 
  \label{fig:s4}
\end{figure*}

\label{sec:bda_2}

\begin{figure*}[t!]
    \centering
    \includegraphics[width=0.85\textwidth]{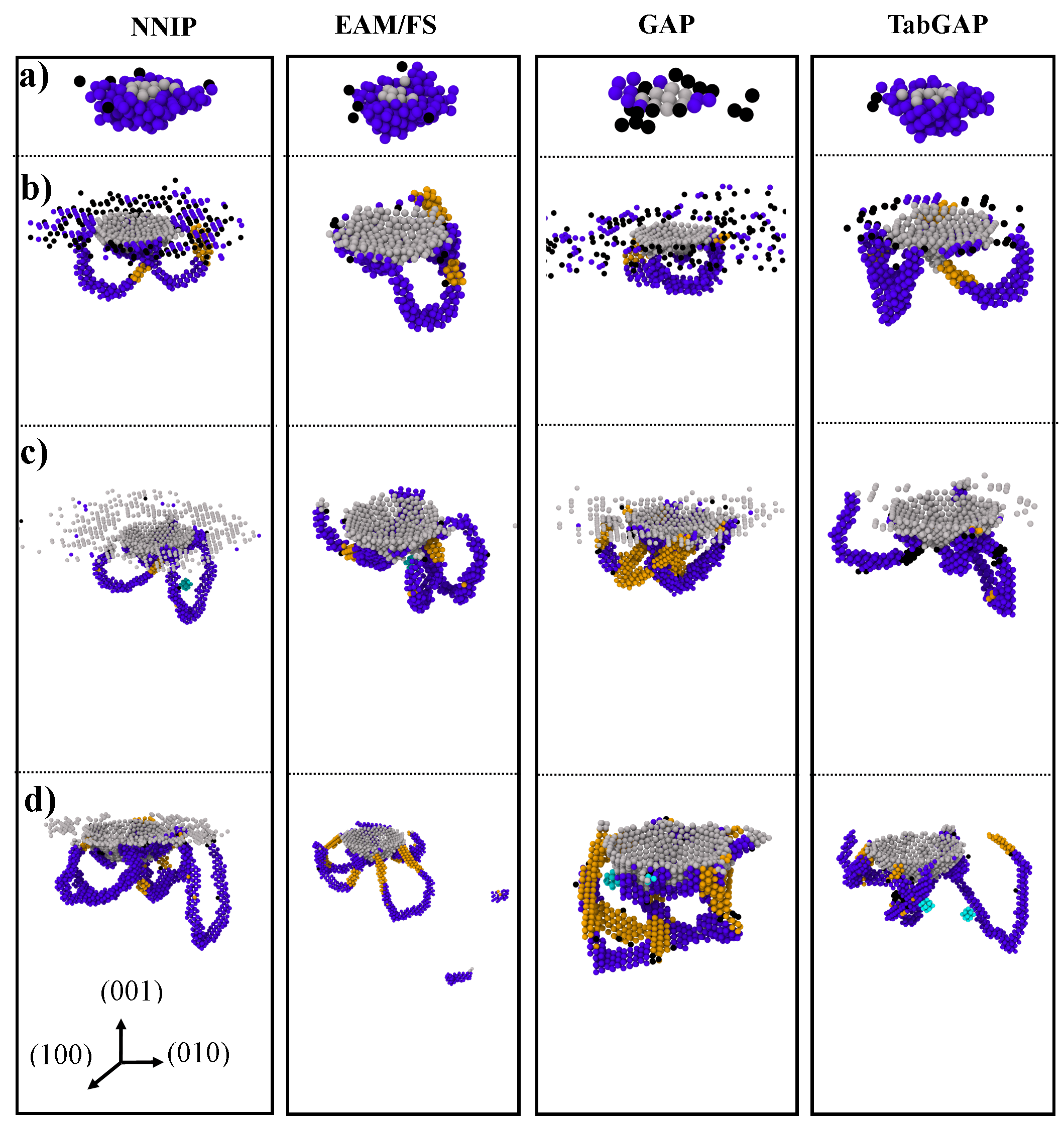}
    \caption{(Color online) Identified defects of indented 
    (001) Mo sample by BDA method at different depths by NNIP, 
    EAM, TabGAP, and GAP approaches. Material defects are 
    depicted using different colors: gray spheres represent 
    surface atoms in direct contact with the indenter tip, blue 
    spheres indicate edge dislocations, light-blue spheres 
    represent atoms in the vicinity of vacancies, yellow spheres 
    depict twin/screw dislocations, and black spheres highlight 
    unidentified defect atoms.
    The nucleation and propagation of edge dislocations on the 
    \{111\} slip 
    family are observed, which then evolve into prismatic loops.
    In addition, identified slip traces and pile-ups are well 
    modeled by NNIP 
    simulations showing the well--known three--fold symmetric 
    rosette depths below 1.40 nm that are 
    formed by [11$\Bar{2}$], [$\Bar{1}$01] and [0$\Bar{1}$1] 
    planes.}
    \label{fig:s5}
\end{figure*}

\begin{figure*}[t!]
    \centering
    \includegraphics[width=0.85\textwidth]{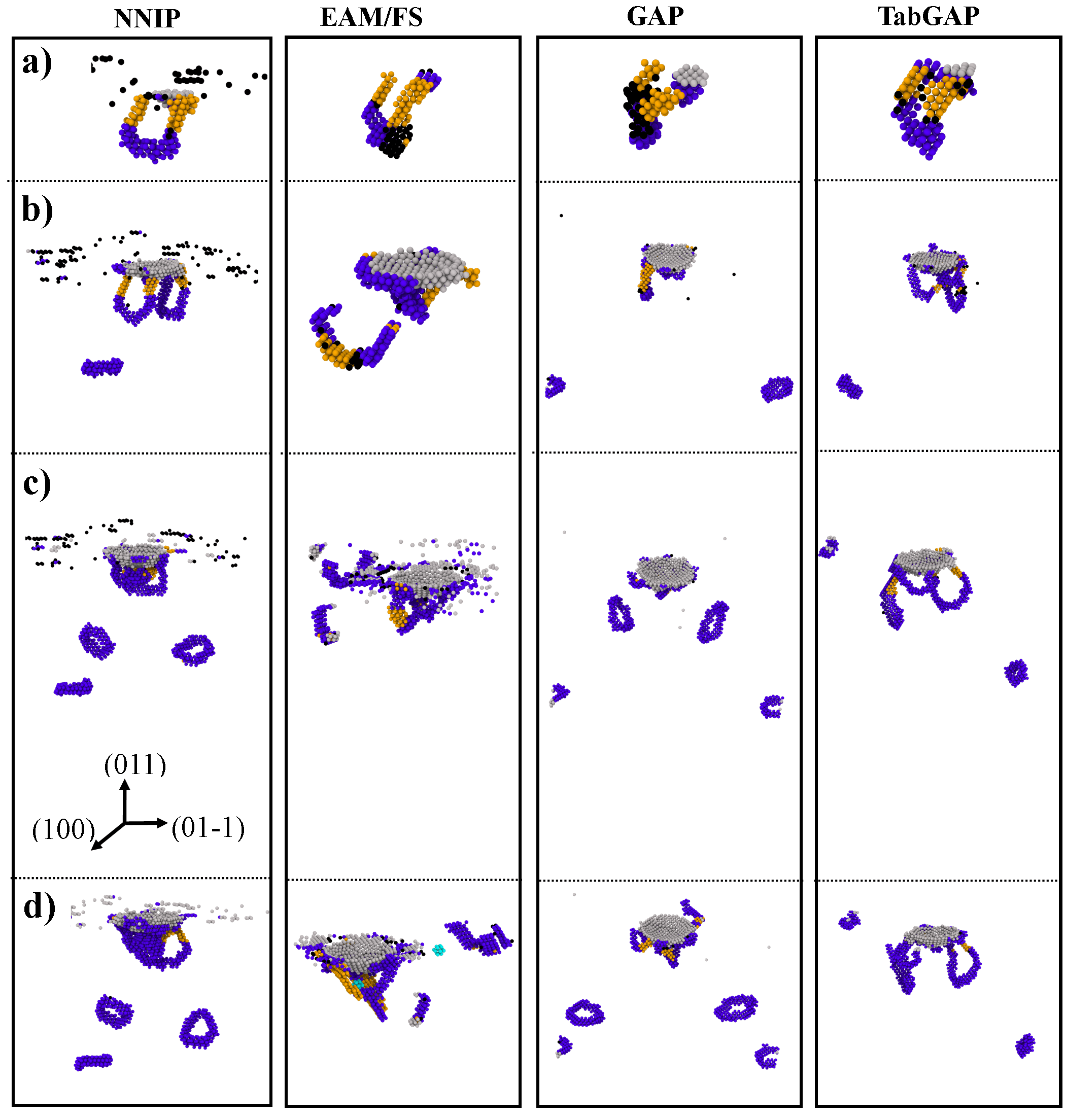}
    \caption{(Color online) Identified defects of indented 
    (011) Mo sample by BDA method at different depths by NNIP, 
    EAM, TabGAP, and GAP approaches. Material defects are 
    depicted using different colors: gray spheres represent 
    surface atoms in direct contact with the indenter tip, blue 
    spheres indicate edge dislocations, light-blue spheres 
    represent atoms in the vicinity of vacancies, yellow spheres 
    depict twin/screw dislocations, and black spheres highlight 
    unidentified defect atoms.
    The nucleation and propagation of edge dislocations on the 
    \{111\} slip 
    family are observed, which then evolve into prismatic loops.
    In addition, identified slip traces and pile-ups are well 
    modeled by NNIP 
    simulations showing the well--known three--fold symmetric 
    rosette depths below 1.40 nm that are 
    formed by [11$\Bar{2}$], [$\Bar{1}$01] and [0$\Bar{1}$1] 
    planes. }
    \label{fig:s6}
\end{figure*}

%% file: main.bbl
\begin{thebibliography}{74}%
\makeatletter
\providecommand \@ifxundefined [1]{%
 \@ifx{#1\undefined}
}%
\providecommand \@ifnum [1]{%
 \ifnum #1\expandafter \@firstoftwo
 \else \expandafter \@secondoftwo
 \fi
}%
\providecommand \@ifx [1]{%
 \ifx #1\expandafter \@firstoftwo
 \else \expandafter \@secondoftwo
 \fi
}%
\providecommand \natexlab [1]{#1}%
\providecommand \enquote  [1]{``#1''}%
\providecommand \bibnamefont  [1]{#1}%
\providecommand \bibfnamefont [1]{#1}%
\providecommand \citenamefont [1]{#1}%
\providecommand \href@noop [0]{\@secondoftwo}%
\providecommand \href [0]{\begingroup \@sanitize@url \@href}%
\providecommand \@href[1]{\@@startlink{#1}\@@href}%
\providecommand \@@href[1]{\endgroup#1\@@endlink}%
\providecommand \@sanitize@url [0]{\catcode `\\12\catcode `\$12\catcode
  `\&12\catcode `\#12\catcode `\^12\catcode `\_12\catcode `\%12\relax}%
\providecommand \@@startlink[1]{}%
\providecommand \@@endlink[0]{}%
\providecommand \url  [0]{\begingroup\@sanitize@url \@url }%
\providecommand \@url [1]{\endgroup\@href {#1}{\urlprefix }}%
\providecommand \urlprefix  [0]{URL }%
\providecommand \Eprint [0]{\href }%
\providecommand \doibase [0]{http://dx.doi.org/}%
\providecommand \selectlanguage [0]{\@gobble}%
\providecommand \bibinfo  [0]{\@secondoftwo}%
\providecommand \bibfield  [0]{\@secondoftwo}%
\providecommand \translation [1]{[#1]}%
\providecommand \BibitemOpen [0]{}%
\providecommand \bibitemStop [0]{}%
\providecommand \bibitemNoStop [0]{.\EOS\space}%
\providecommand \EOS [0]{\spacefactor3000\relax}%
\providecommand \BibitemShut  [1]{\csname bibitem#1\endcsname}%
\let\auto@bib@innerbib\@empty
\bibitem [{\citenamefont {Kim}\ and\ \citenamefont
  {Greer}(2009)}]{KIM20095245}%
  \BibitemOpen
  \bibfield  {author} {\bibinfo {author} {\bibfnamefont {J.-Y.}\ \bibnamefont
  {Kim}}\ and\ \bibinfo {author} {\bibfnamefont {J.~R.}\ \bibnamefont
  {Greer}},\ }\href {\doibase https://doi.org/10.1016/j.actamat.2009.07.027}
  {\bibfield  {journal} {\bibinfo  {journal} {Acta Materialia}\ }\textbf
  {\bibinfo {volume} {57}},\ \bibinfo {pages} {5245} (\bibinfo {year}
  {2009})}\BibitemShut {NoStop}%
\bibitem [{\citenamefont {Kim}\ \emph {et~al.}(2012)\citenamefont {Kim},
  \citenamefont {Jang},\ and\ \citenamefont {Greer}}]{KIM201246}%
  \BibitemOpen
  \bibfield  {author} {\bibinfo {author} {\bibfnamefont {J.-Y.}\ \bibnamefont
  {Kim}}, \bibinfo {author} {\bibfnamefont {D.}~\bibnamefont {Jang}}, \ and\
  \bibinfo {author} {\bibfnamefont {J.~R.}\ \bibnamefont {Greer}},\ }\href
  {\doibase https://doi.org/10.1016/j.ijplas.2011.05.015} {\bibfield  {journal}
  {\bibinfo  {journal} {International Journal of Plasticity}\ }\textbf
  {\bibinfo {volume} {28}},\ \bibinfo {pages} {46} (\bibinfo {year}
  {2012})}\BibitemShut {NoStop}%
\bibitem [{\citenamefont {Motamedi}\ \emph {et~al.}(2020)\citenamefont
  {Motamedi}, \citenamefont {Naghdi},\ and\ \citenamefont
  {Jalali}}]{amirtensile}%
  \BibitemOpen
  \bibfield  {author} {\bibinfo {author} {\bibfnamefont {M.}~\bibnamefont
  {Motamedi}}, \bibinfo {author} {\bibfnamefont {A.}~\bibnamefont {Naghdi}}, \
  and\ \bibinfo {author} {\bibfnamefont {S.}~\bibnamefont {Jalali}},\ }\href
  {\doibase 10.1177/0954406219878760} {\bibfield  {journal} {\bibinfo
  {journal} {Proceedings of the Institution of Mechanical Engineers, Part C:
  Journal of Mechanical Engineering Science}\ }\textbf {\bibinfo {volume}
  {234}},\ \bibinfo {pages} {635} (\bibinfo {year} {2020})},\ \Eprint
  {http://arxiv.org/abs/https://doi.org/10.1177/0954406219878760}
  {https://doi.org/10.1177/0954406219878760} \BibitemShut {NoStop}%
\bibitem [{\citenamefont {Schuh}(2006)}]{SCHUH200632}%
  \BibitemOpen
  \bibfield  {author} {\bibinfo {author} {\bibfnamefont {C.~A.}\ \bibnamefont
  {Schuh}},\ }\href {\doibase https://doi.org/10.1016/S1369-7021(06)71495-X}
  {\bibfield  {journal} {\bibinfo  {journal} {Materials Today}\ }\textbf
  {\bibinfo {volume} {9}},\ \bibinfo {pages} {32} (\bibinfo {year}
  {2006})}\BibitemShut {NoStop}%
\bibitem [{\citenamefont {Varillas}\ \emph {et~al.}(2017)\citenamefont
  {Varillas}, \citenamefont {Ocenasek}, \citenamefont {Torner},\ and\
  \citenamefont {Alcala}}]{VARILLAS2017431}%
  \BibitemOpen
  \bibfield  {author} {\bibinfo {author} {\bibfnamefont {J.}~\bibnamefont
  {Varillas}}, \bibinfo {author} {\bibfnamefont {J.}~\bibnamefont {Ocenasek}},
  \bibinfo {author} {\bibfnamefont {J.}~\bibnamefont {Torner}}, \ and\ \bibinfo
  {author} {\bibfnamefont {J.}~\bibnamefont {Alcala}},\ }\href {\doibase
  https://doi.org/10.1016/j.actamat.2016.11.067} {\bibfield  {journal}
  {\bibinfo  {journal} {Acta Materialia}\ }\textbf {\bibinfo {volume} {125}},\
  \bibinfo {pages} {431} (\bibinfo {year} {2017})}\BibitemShut {NoStop}%
\bibitem [{\citenamefont {Kurpaska}\ \emph {et~al.}(2022)\citenamefont
  {Kurpaska}, \citenamefont {Dominguez-Gutierrez}, \citenamefont {Zhang},
  \citenamefont {Mulewska}, \citenamefont {Bei}, \citenamefont {Weber},
  \citenamefont {Kosińska}, \citenamefont {Chrominski}, \citenamefont
  {Jozwik}, \citenamefont {Alvarez-Donado}, \citenamefont {Papanikolaou},
  \citenamefont {Jagielski},\ and\ \citenamefont {Alava}}]{KURPASKA2022110639}%
  \BibitemOpen
  \bibfield  {author} {\bibinfo {author} {\bibfnamefont {L.}~\bibnamefont
  {Kurpaska}}, \bibinfo {author} {\bibfnamefont {F.}~\bibnamefont
  {Dominguez-Gutierrez}}, \bibinfo {author} {\bibfnamefont {Y.}~\bibnamefont
  {Zhang}}, \bibinfo {author} {\bibfnamefont {K.}~\bibnamefont {Mulewska}},
  \bibinfo {author} {\bibfnamefont {H.}~\bibnamefont {Bei}}, \bibinfo {author}
  {\bibfnamefont {W.}~\bibnamefont {Weber}}, \bibinfo {author} {\bibfnamefont
  {A.}~\bibnamefont {Kosińska}}, \bibinfo {author} {\bibfnamefont
  {W.}~\bibnamefont {Chrominski}}, \bibinfo {author} {\bibfnamefont
  {I.}~\bibnamefont {Jozwik}}, \bibinfo {author} {\bibfnamefont
  {R.}~\bibnamefont {Alvarez-Donado}}, \bibinfo {author} {\bibfnamefont
  {S.}~\bibnamefont {Papanikolaou}}, \bibinfo {author} {\bibfnamefont
  {J.}~\bibnamefont {Jagielski}}, \ and\ \bibinfo {author} {\bibfnamefont
  {M.}~\bibnamefont {Alava}},\ }\href {\doibase
  https://doi.org/10.1016/j.matdes.2022.110639} {\bibfield  {journal} {\bibinfo
   {journal} {Materials \& Design}\ }\textbf {\bibinfo {volume} {217}},\
  \bibinfo {pages} {110639} (\bibinfo {year} {2022})}\BibitemShut {NoStop}%
\bibitem [{\citenamefont {Pathak}\ and\ \citenamefont
  {Kalidindi}(2015)}]{PATHAK20151}%
  \BibitemOpen
  \bibfield  {author} {\bibinfo {author} {\bibfnamefont {S.}~\bibnamefont
  {Pathak}}\ and\ \bibinfo {author} {\bibfnamefont {S.~R.}\ \bibnamefont
  {Kalidindi}},\ }\href@noop {} {\bibfield  {journal} {\bibinfo  {journal}
  {Materials Science and Engineering: R: Reports}\ }\textbf {\bibinfo {volume}
  {91}},\ \bibinfo {pages} {1} (\bibinfo {year} {2015})}\BibitemShut {NoStop}%
\bibitem [{\citenamefont {Voyiadjis}\ and\ \citenamefont
  {Yaghoobi}(2017)}]{cryst7100321}%
  \BibitemOpen
  \bibfield  {author} {\bibinfo {author} {\bibfnamefont {G.~Z.}\ \bibnamefont
  {Voyiadjis}}\ and\ \bibinfo {author} {\bibfnamefont {M.}~\bibnamefont
  {Yaghoobi}},\ }\href@noop {} {\bibfield  {journal} {\bibinfo  {journal}
  {Crystals}\ }\textbf {\bibinfo {volume} {7}},\ \bibinfo {pages} {321}
  (\bibinfo {year} {2017})}\BibitemShut {NoStop}%
\bibitem [{\citenamefont {Remington}\ \emph {et~al.}(2014)\citenamefont
  {Remington}, \citenamefont {Ruestes}, \citenamefont {Bringa}, \citenamefont
  {Remington}, \citenamefont {Lu}, \citenamefont {Kad},\ and\ \citenamefont
  {Meyers}}]{REMINGTON2014378}%
  \BibitemOpen
  \bibfield  {author} {\bibinfo {author} {\bibfnamefont {T.}~\bibnamefont
  {Remington}}, \bibinfo {author} {\bibfnamefont {C.}~\bibnamefont {Ruestes}},
  \bibinfo {author} {\bibfnamefont {E.}~\bibnamefont {Bringa}}, \bibinfo
  {author} {\bibfnamefont {B.}~\bibnamefont {Remington}}, \bibinfo {author}
  {\bibfnamefont {C.}~\bibnamefont {Lu}}, \bibinfo {author} {\bibfnamefont
  {B.}~\bibnamefont {Kad}}, \ and\ \bibinfo {author} {\bibfnamefont
  {M.}~\bibnamefont {Meyers}},\ }\href {\doibase
  https://doi.org/10.1016/j.actamat.2014.06.058} {\bibfield  {journal}
  {\bibinfo  {journal} {Acta Materialia}\ }\textbf {\bibinfo {volume} {78}},\
  \bibinfo {pages} {378} (\bibinfo {year} {2014})}\BibitemShut {NoStop}%
\bibitem [{\citenamefont {Gagel}\ \emph {et~al.}(2016)\citenamefont {Gagel},
  \citenamefont {Weygand},\ and\ \citenamefont {Gumbsch}}]{GAGEL2016399}%
  \BibitemOpen
  \bibfield  {author} {\bibinfo {author} {\bibfnamefont {J.}~\bibnamefont
  {Gagel}}, \bibinfo {author} {\bibfnamefont {D.}~\bibnamefont {Weygand}}, \
  and\ \bibinfo {author} {\bibfnamefont {P.}~\bibnamefont {Gumbsch}},\
  }\href@noop {} {\bibfield  {journal} {\bibinfo  {journal} {Acta Materialia}\
  }\textbf {\bibinfo {volume} {111}},\ \bibinfo {pages} {399} (\bibinfo {year}
  {2016})}\BibitemShut {NoStop}%
\bibitem [{\citenamefont {Naghdi}\ \emph {et~al.}(2022)\citenamefont {Naghdi},
  \citenamefont {Dominguez-Gutierrez}, \citenamefont {Huo}, \citenamefont
  {Karimi},\ and\ \citenamefont {Papanikolaou}}]{naghdi2022dynamic}%
  \BibitemOpen
  \bibfield  {author} {\bibinfo {author} {\bibfnamefont {A.}~\bibnamefont
  {Naghdi}}, \bibinfo {author} {\bibfnamefont {F.~J.}\ \bibnamefont
  {Dominguez-Gutierrez}}, \bibinfo {author} {\bibfnamefont {W.~Y.}\
  \bibnamefont {Huo}}, \bibinfo {author} {\bibfnamefont {K.}~\bibnamefont
  {Karimi}}, \ and\ \bibinfo {author} {\bibfnamefont {S.}~\bibnamefont
  {Papanikolaou}},\ }\href@noop {} {\enquote {\bibinfo {title} {Dynamic
  nanoindentation and short-range order in equiatomic nicocr medium entropy
  alloy lead to novel density wave ordering},}\ } (\bibinfo {year} {2022}),\
  \Eprint {http://arxiv.org/abs/2211.05436} {arXiv:2211.05436
  [cond-mat.mtrl-sci]} \BibitemShut {NoStop}%
\bibitem [{\citenamefont {Taneike}\ \emph {et~al.}(2003)\citenamefont
  {Taneike}, \citenamefont {Abe},\ and\ \citenamefont {Sawada}}]{Taneike2003}%
  \BibitemOpen
  \bibfield  {author} {\bibinfo {author} {\bibfnamefont {M.}~\bibnamefont
  {Taneike}}, \bibinfo {author} {\bibfnamefont {F.}~\bibnamefont {Abe}}, \ and\
  \bibinfo {author} {\bibfnamefont {K.}~\bibnamefont {Sawada}},\ }\href
  {\doibase 10.1038/nature01740} {\bibfield  {journal} {\bibinfo  {journal}
  {Nature}\ }\textbf {\bibinfo {volume} {424}},\ \bibinfo {pages} {294}
  (\bibinfo {year} {2003})}\BibitemShut {NoStop}%
\bibitem [{\citenamefont {Haque}\ and\ \citenamefont {Saif}(2002)}]{Haque2002}%
  \BibitemOpen
  \bibfield  {author} {\bibinfo {author} {\bibfnamefont {M.~A.}\ \bibnamefont
  {Haque}}\ and\ \bibinfo {author} {\bibfnamefont {M.~T.~A.}\ \bibnamefont
  {Saif}},\ }\href {\doibase 10.1007/BF02411059} {\bibfield  {journal}
  {\bibinfo  {journal} {Experimental Mechanics}\ }\textbf {\bibinfo {volume}
  {42}},\ \bibinfo {pages} {123} (\bibinfo {year} {2002})}\BibitemShut
  {NoStop}%
\bibitem [{\citenamefont {De~Hosson}\ \emph {et~al.}(2006)\citenamefont
  {De~Hosson}, \citenamefont {Soer}, \citenamefont {Minor}, \citenamefont
  {Shan}, \citenamefont {Stach}, \citenamefont {Syed~Asif},\ and\ \citenamefont
  {Warren}}]{DeHosson2006}%
  \BibitemOpen
  \bibfield  {author} {\bibinfo {author} {\bibfnamefont {J.~T.~M.}\
  \bibnamefont {De~Hosson}}, \bibinfo {author} {\bibfnamefont {W.~A.}\
  \bibnamefont {Soer}}, \bibinfo {author} {\bibfnamefont {A.~M.}\ \bibnamefont
  {Minor}}, \bibinfo {author} {\bibfnamefont {Z.}~\bibnamefont {Shan}},
  \bibinfo {author} {\bibfnamefont {E.~A.}\ \bibnamefont {Stach}}, \bibinfo
  {author} {\bibfnamefont {S.~A.}\ \bibnamefont {Syed~Asif}}, \ and\ \bibinfo
  {author} {\bibfnamefont {O.~L.}\ \bibnamefont {Warren}},\ }\href {\doibase
  10.1007/s10853-006-0472-2} {\bibfield  {journal} {\bibinfo  {journal}
  {Journal of Materials Science}\ }\textbf {\bibinfo {volume} {41}},\ \bibinfo
  {pages} {7704} (\bibinfo {year} {2006})}\BibitemShut {NoStop}%
\bibitem [{\citenamefont {Durst}\ \emph {et~al.}(2005)\citenamefont {Durst},
  \citenamefont {Backes},\ and\ \citenamefont {Göken}}]{durst}%
  \BibitemOpen
  \bibfield  {author} {\bibinfo {author} {\bibfnamefont {K.}~\bibnamefont
  {Durst}}, \bibinfo {author} {\bibfnamefont {B.}~\bibnamefont {Backes}}, \
  and\ \bibinfo {author} {\bibfnamefont {M.}~\bibnamefont {Göken}},\ }\href
  {\doibase https://doi.org/10.1016/j.scriptamat.2005.02.009} {\bibfield
  {journal} {\bibinfo  {journal} {Scripta Materialia}\ }\textbf {\bibinfo
  {volume} {52}},\ \bibinfo {pages} {1093} (\bibinfo {year}
  {2005})}\BibitemShut {NoStop}%
\bibitem [{\citenamefont {Maier-Kiener}\ \emph {et~al.}(2017)\citenamefont
  {Maier-Kiener}, \citenamefont {Verena}, \citenamefont {Durst},\ and\
  \citenamefont {Karsten}}]{MaierKiener2017}%
  \BibitemOpen
  \bibfield  {author} {\bibinfo {author} {\bibnamefont {Maier-Kiener}},
  \bibinfo {author} {\bibnamefont {Verena}}, \bibinfo {author} {\bibnamefont
  {Durst}}, \ and\ \bibinfo {author} {\bibnamefont {Karsten}},\ }\href
  {\doibase 10.1007/s11837-017-2536-y} {\bibfield  {journal} {\bibinfo
  {journal} {JOM}\ }\textbf {\bibinfo {volume} {69}},\ \bibinfo {pages} {2246}
  (\bibinfo {year} {2017})}\BibitemShut {NoStop}%
\bibitem [{\citenamefont {Casals}\ \emph {et~al.}(2007)\citenamefont {Casals},
  \citenamefont {Očenašek},\ and\ \citenamefont {Alcala}}]{CASALS200755}%
  \BibitemOpen
  \bibfield  {author} {\bibinfo {author} {\bibfnamefont {O.}~\bibnamefont
  {Casals}}, \bibinfo {author} {\bibfnamefont {J.}~\bibnamefont {Očenašek}},
  \ and\ \bibinfo {author} {\bibfnamefont {J.}~\bibnamefont {Alcala}},\ }\href
  {\doibase https://doi.org/10.1016/j.actamat.2006.07.018} {\bibfield
  {journal} {\bibinfo  {journal} {Acta Materialia}\ }\textbf {\bibinfo {volume}
  {55}},\ \bibinfo {pages} {55} (\bibinfo {year} {2007})}\BibitemShut {NoStop}%
\bibitem [{\citenamefont {Smith}\ \emph {et~al.}(2003)\citenamefont {Smith},
  \citenamefont {Christopher}, \citenamefont {Kenny}, \citenamefont {Richter},\
  and\ \citenamefont {Wolf}}]{PhysRevB.67.245405}%
  \BibitemOpen
  \bibfield  {author} {\bibinfo {author} {\bibfnamefont {R.}~\bibnamefont
  {Smith}}, \bibinfo {author} {\bibfnamefont {D.}~\bibnamefont {Christopher}},
  \bibinfo {author} {\bibfnamefont {S.~D.}\ \bibnamefont {Kenny}}, \bibinfo
  {author} {\bibfnamefont {A.}~\bibnamefont {Richter}}, \ and\ \bibinfo
  {author} {\bibfnamefont {B.}~\bibnamefont {Wolf}},\ }\href {\doibase
  10.1103/PhysRevB.67.245405} {\bibfield  {journal} {\bibinfo  {journal} {Phys.
  Rev. B}\ }\textbf {\bibinfo {volume} {67}},\ \bibinfo {pages} {245405}
  (\bibinfo {year} {2003})}\BibitemShut {NoStop}%
\bibitem [{\citenamefont {Swaddiwudhipong}\ \emph {et~al.}(2006)\citenamefont
  {Swaddiwudhipong}, \citenamefont {Tho}, \citenamefont {Hua},\ and\
  \citenamefont {Liu}}]{SWADDIWUDHIPONG20061117}%
  \BibitemOpen
  \bibfield  {author} {\bibinfo {author} {\bibfnamefont {S.}~\bibnamefont
  {Swaddiwudhipong}}, \bibinfo {author} {\bibfnamefont {K.}~\bibnamefont
  {Tho}}, \bibinfo {author} {\bibfnamefont {J.}~\bibnamefont {Hua}}, \ and\
  \bibinfo {author} {\bibfnamefont {Z.}~\bibnamefont {Liu}},\ }\href {\doibase
  https://doi.org/10.1016/j.ijsolstr.2005.05.026} {\bibfield  {journal}
  {\bibinfo  {journal} {International Journal of Solids and Structures}\
  }\textbf {\bibinfo {volume} {43}},\ \bibinfo {pages} {1117} (\bibinfo {year}
  {2006})}\BibitemShut {NoStop}%
\bibitem [{\citenamefont {Dominguez-Gutierrez}\ \emph
  {et~al.}(2021)\citenamefont {Dominguez-Gutierrez}, \citenamefont
  {Papanikolaou}, \citenamefont {Esfandiarpour}, \citenamefont {Sobkowicz},\
  and\ \citenamefont {Alava}}]{DOMINGUEZGUTIERREZ2021141912}%
  \BibitemOpen
  \bibfield  {author} {\bibinfo {author} {\bibfnamefont {F.}~\bibnamefont
  {Dominguez-Gutierrez}}, \bibinfo {author} {\bibfnamefont {S.}~\bibnamefont
  {Papanikolaou}}, \bibinfo {author} {\bibfnamefont {A.}~\bibnamefont
  {Esfandiarpour}}, \bibinfo {author} {\bibfnamefont {P.}~\bibnamefont
  {Sobkowicz}}, \ and\ \bibinfo {author} {\bibfnamefont {M.}~\bibnamefont
  {Alava}},\ }\href {\doibase https://doi.org/10.1016/j.msea.2021.141912}
  {\bibfield  {journal} {\bibinfo  {journal} {Materials Science and
  Engineering: A}\ }\textbf {\bibinfo {volume} {826}},\ \bibinfo {pages}
  {141912} (\bibinfo {year} {2021})}\BibitemShut {NoStop}%
\bibitem [{\citenamefont {Kramer}\ \emph {et~al.}(2001)\citenamefont {Kramer},
  \citenamefont {Yoder},\ and\ \citenamefont {Gerberich}}]{article}%
  \BibitemOpen
  \bibfield  {author} {\bibinfo {author} {\bibfnamefont {D.}~\bibnamefont
  {Kramer}}, \bibinfo {author} {\bibfnamefont {K.}~\bibnamefont {Yoder}}, \
  and\ \bibinfo {author} {\bibfnamefont {W.}~\bibnamefont {Gerberich}},\ }\href
  {\doibase 10.1080/01418610108216651} {\bibfield  {journal} {\bibinfo
  {journal} {Philosophical Magazine A-physics of Condensed Matter Structure
  Defects and Mechanical Properties - PHIL MAG A}\ }\textbf {\bibinfo {volume}
  {81}},\ \bibinfo {pages} {2033} (\bibinfo {year} {2001})}\BibitemShut
  {NoStop}%
\bibitem [{\citenamefont {Biener}\ \emph {et~al.}(2007)\citenamefont {Biener},
  \citenamefont {Biener}, \citenamefont {Hodge},\ and\ \citenamefont
  {Hamza}}]{PhysRevB.76.165422}%
  \BibitemOpen
  \bibfield  {author} {\bibinfo {author} {\bibfnamefont {M.~M.}\ \bibnamefont
  {Biener}}, \bibinfo {author} {\bibfnamefont {J.}~\bibnamefont {Biener}},
  \bibinfo {author} {\bibfnamefont {A.~M.}\ \bibnamefont {Hodge}}, \ and\
  \bibinfo {author} {\bibfnamefont {A.~V.}\ \bibnamefont {Hamza}},\ }\href
  {\doibase 10.1103/PhysRevB.76.165422} {\bibfield  {journal} {\bibinfo
  {journal} {Phys. Rev. B}\ }\textbf {\bibinfo {volume} {76}},\ \bibinfo
  {pages} {165422} (\bibinfo {year} {2007})}\BibitemShut {NoStop}%
\bibitem [{\citenamefont {Voyiadjis}\ and\ \citenamefont
  {Faghihi}(2012)}]{VOYIADJIS2012205}%
  \BibitemOpen
  \bibfield  {author} {\bibinfo {author} {\bibfnamefont {G.~Z.}\ \bibnamefont
  {Voyiadjis}}\ and\ \bibinfo {author} {\bibfnamefont {D.}~\bibnamefont
  {Faghihi}},\ }\href {\doibase https://doi.org/10.1016/j.piutam.2012.03.014}
  {\bibfield  {journal} {\bibinfo  {journal} {Procedia IUTAM}\ }\textbf
  {\bibinfo {volume} {3}},\ \bibinfo {pages} {205} (\bibinfo {year} {2012})},\
  \bibinfo {note} {iUTAM Symposium on Linking Scales in Computations: From
  Microstructure to Macro-scale Properties}\BibitemShut {NoStop}%
\bibitem [{\citenamefont {Plummer}(2019)}]{plummer}%
  \BibitemOpen
  \bibfield  {author} {\bibinfo {author} {\bibfnamefont {K.~P.~E.}\
  \bibnamefont {Plummer}},\ }\href@noop {} {\bibfield  {journal} {\bibinfo
  {journal} {University of Oxford}\ } (\bibinfo {year} {2019})}\BibitemShut
  {NoStop}%
\bibitem [{\citenamefont {Frydrych}\ \emph {et~al.}(2023)\citenamefont
  {Frydrych}, \citenamefont {Dominguez-Gutierrez}, \citenamefont {Alava},\ and\
  \citenamefont {Papanikolaou}}]{FRYDRYCH2023104644}%
  \BibitemOpen
  \bibfield  {author} {\bibinfo {author} {\bibfnamefont {K.}~\bibnamefont
  {Frydrych}}, \bibinfo {author} {\bibfnamefont {F.}~\bibnamefont
  {Dominguez-Gutierrez}}, \bibinfo {author} {\bibfnamefont {M.}~\bibnamefont
  {Alava}}, \ and\ \bibinfo {author} {\bibfnamefont {S.}~\bibnamefont
  {Papanikolaou}},\ }\href {\doibase
  https://doi.org/10.1016/j.mechmat.2023.104644} {\bibfield  {journal}
  {\bibinfo  {journal} {Mechanics of Materials}\ }\textbf {\bibinfo {volume}
  {181}},\ \bibinfo {pages} {104644} (\bibinfo {year} {2023})}\BibitemShut
  {NoStop}%
\bibitem [{\citenamefont {Knapp}\ \emph {et~al.}(1999)\citenamefont {Knapp},
  \citenamefont {Follstaedt}, \citenamefont {Myers}, \citenamefont {Barbour},\
  and\ \citenamefont {Friedmann}}]{10.1063/1.369178}%
  \BibitemOpen
  \bibfield  {author} {\bibinfo {author} {\bibfnamefont {J.~A.}\ \bibnamefont
  {Knapp}}, \bibinfo {author} {\bibfnamefont {D.~M.}\ \bibnamefont
  {Follstaedt}}, \bibinfo {author} {\bibfnamefont {S.~M.}\ \bibnamefont
  {Myers}}, \bibinfo {author} {\bibfnamefont {J.~C.}\ \bibnamefont {Barbour}},
  \ and\ \bibinfo {author} {\bibfnamefont {T.~A.}\ \bibnamefont {Friedmann}},\
  }\href {\doibase 10.1063/1.369178} {\bibfield  {journal} {\bibinfo  {journal}
  {Journal of Applied Physics}\ }\textbf {\bibinfo {volume} {85}},\ \bibinfo
  {pages} {1460} (\bibinfo {year} {1999})},\ \Eprint
  {http://arxiv.org/abs/https://pubs.aip.org/aip/jap/article-pdf/85/3/1460/10597542/1460\_1\_online.pdf}
  {https://pubs.aip.org/aip/jap/article-pdf/85/3/1460/10597542/1460\_1\_online.pdf}
  \BibitemShut {NoStop}%
\bibitem [{\citenamefont {Lichinchi}\ \emph {et~al.}(1998)\citenamefont
  {Lichinchi}, \citenamefont {Lenardi}, \citenamefont {Haupt},\ and\
  \citenamefont {Vitali}}]{LICHINCHI1998240}%
  \BibitemOpen
  \bibfield  {author} {\bibinfo {author} {\bibfnamefont {M.}~\bibnamefont
  {Lichinchi}}, \bibinfo {author} {\bibfnamefont {C.}~\bibnamefont {Lenardi}},
  \bibinfo {author} {\bibfnamefont {J.}~\bibnamefont {Haupt}}, \ and\ \bibinfo
  {author} {\bibfnamefont {R.}~\bibnamefont {Vitali}},\ }\href {\doibase
  https://doi.org/10.1016/S0040-6090(97)00739-6} {\bibfield  {journal}
  {\bibinfo  {journal} {Thin Solid Films}\ }\textbf {\bibinfo {volume} {312}},\
  \bibinfo {pages} {240} (\bibinfo {year} {1998})}\BibitemShut {NoStop}%
\bibitem [{\citenamefont {Zhou}\ \emph {et~al.}(2010)\citenamefont {Zhou},
  \citenamefont {Biner},\ and\ \citenamefont {LeSar}}]{ZHOU20101565}%
  \BibitemOpen
  \bibfield  {author} {\bibinfo {author} {\bibfnamefont {C.}~\bibnamefont
  {Zhou}}, \bibinfo {author} {\bibfnamefont {S.~B.}\ \bibnamefont {Biner}}, \
  and\ \bibinfo {author} {\bibfnamefont {R.}~\bibnamefont {LeSar}},\ }\href
  {\doibase https://doi.org/10.1016/j.actamat.2009.11.001} {\bibfield
  {journal} {\bibinfo  {journal} {Acta Materialia}\ }\textbf {\bibinfo {volume}
  {58}},\ \bibinfo {pages} {1565} (\bibinfo {year} {2010})}\BibitemShut
  {NoStop}%
\bibitem [{\citenamefont {Motz}\ \emph {et~al.}(2008)\citenamefont {Motz},
  \citenamefont {Weygand}, \citenamefont {Senger},\ and\ \citenamefont
  {Gumbsch}}]{MOTZ20081942}%
  \BibitemOpen
  \bibfield  {author} {\bibinfo {author} {\bibfnamefont {C.}~\bibnamefont
  {Motz}}, \bibinfo {author} {\bibfnamefont {D.}~\bibnamefont {Weygand}},
  \bibinfo {author} {\bibfnamefont {J.}~\bibnamefont {Senger}}, \ and\ \bibinfo
  {author} {\bibfnamefont {P.}~\bibnamefont {Gumbsch}},\ }\href {\doibase
  https://doi.org/10.1016/j.actamat.2007.12.053} {\bibfield  {journal}
  {\bibinfo  {journal} {Acta Materialia}\ }\textbf {\bibinfo {volume} {56}},\
  \bibinfo {pages} {1942} (\bibinfo {year} {2008})}\BibitemShut {NoStop}%
\bibitem [{\citenamefont {Song}\ \emph {et~al.}(2019)\citenamefont {Song},
  \citenamefont {Yavas}, \citenamefont {der Giessen},\ and\ \citenamefont
  {Papanikolaou}}]{SONG2019332}%
  \BibitemOpen
  \bibfield  {author} {\bibinfo {author} {\bibfnamefont {H.}~\bibnamefont
  {Song}}, \bibinfo {author} {\bibfnamefont {H.}~\bibnamefont {Yavas}},
  \bibinfo {author} {\bibfnamefont {E.~V.}\ \bibnamefont {der Giessen}}, \ and\
  \bibinfo {author} {\bibfnamefont {S.}~\bibnamefont {Papanikolaou}},\ }\href
  {\doibase https://doi.org/10.1016/j.jmps.2018.09.005} {\bibfield  {journal}
  {\bibinfo  {journal} {Journal of the Mechanics and Physics of Solids}\
  }\textbf {\bibinfo {volume} {123}},\ \bibinfo {pages} {332} (\bibinfo {year}
  {2019})},\ \bibinfo {note} {the N.A. Fleck 60th Anniversary
  Volume}\BibitemShut {NoStop}%
\bibitem [{\citenamefont {Xu}\ \emph {et~al.}(2024)\citenamefont {Xu},
  \citenamefont {Zaborowska}, \citenamefont {Mulewska}, \citenamefont {Huo},
  \citenamefont {Karimi}, \citenamefont {Domínguez-Gutiérrez}, \citenamefont
  {Łukasz Kurpaska}, \citenamefont {Alava},\ and\ \citenamefont
  {Papanikolaou}}]{XU2024112733}%
  \BibitemOpen
  \bibfield  {author} {\bibinfo {author} {\bibfnamefont {Q.}~\bibnamefont
  {Xu}}, \bibinfo {author} {\bibfnamefont {A.}~\bibnamefont {Zaborowska}},
  \bibinfo {author} {\bibfnamefont {K.}~\bibnamefont {Mulewska}}, \bibinfo
  {author} {\bibfnamefont {W.}~\bibnamefont {Huo}}, \bibinfo {author}
  {\bibfnamefont {K.}~\bibnamefont {Karimi}}, \bibinfo {author} {\bibfnamefont
  {F.~J.}\ \bibnamefont {Domínguez-Gutiérrez}}, \bibinfo {author}
  {\bibnamefont {Łukasz Kurpaska}}, \bibinfo {author} {\bibfnamefont {M.~J.}\
  \bibnamefont {Alava}}, \ and\ \bibinfo {author} {\bibfnamefont
  {S.}~\bibnamefont {Papanikolaou}},\ }\href {\doibase
  https://doi.org/10.1016/j.vacuum.2023.112733} {\bibfield  {journal} {\bibinfo
   {journal} {Vacuum}\ }\textbf {\bibinfo {volume} {219}},\ \bibinfo {pages}
  {112733} (\bibinfo {year} {2024})}\BibitemShut {NoStop}%
\bibitem [{\citenamefont {Dom\'{\i}nguez-Guti\'errez}\ \emph
  {et~al.}(2023)\citenamefont {Dom\'{\i}nguez-Guti\'errez}, \citenamefont
  {Grigorev}, \citenamefont {Naghdi}, \citenamefont {Byggm\"astar},
  \citenamefont {Wei}, \citenamefont {Swinburne}, \citenamefont
  {Papanikolaou},\ and\ \citenamefont {Alava}}]{PhysRevMaterials.7.043603}%
  \BibitemOpen
  \bibfield  {author} {\bibinfo {author} {\bibfnamefont {F.~J.}\ \bibnamefont
  {Dom\'{\i}nguez-Guti\'errez}}, \bibinfo {author} {\bibfnamefont
  {P.}~\bibnamefont {Grigorev}}, \bibinfo {author} {\bibfnamefont
  {A.}~\bibnamefont {Naghdi}}, \bibinfo {author} {\bibfnamefont
  {J.}~\bibnamefont {Byggm\"astar}}, \bibinfo {author} {\bibfnamefont {G.~Y.}\
  \bibnamefont {Wei}}, \bibinfo {author} {\bibfnamefont {T.~D.}\ \bibnamefont
  {Swinburne}}, \bibinfo {author} {\bibfnamefont {S.}~\bibnamefont
  {Papanikolaou}}, \ and\ \bibinfo {author} {\bibfnamefont {M.~J.}\
  \bibnamefont {Alava}},\ }\href {\doibase 10.1103/PhysRevMaterials.7.043603}
  {\bibfield  {journal} {\bibinfo  {journal} {Phys. Rev. Mater.}\ }\textbf
  {\bibinfo {volume} {7}},\ \bibinfo {pages} {043603} (\bibinfo {year}
  {2023})}\BibitemShut {NoStop}%
\bibitem [{\citenamefont {Naghdi}\ \emph {et~al.}(2023)\citenamefont {Naghdi},
  \citenamefont {Karimi}, \citenamefont {Poisvert}, \citenamefont
  {Esfandiarpour}, \citenamefont {Alvarez}, \citenamefont {Sobkowicz},
  \citenamefont {Alava},\ and\ \citenamefont
  {Papanikolaou}}]{PhysRevB.107.094109}%
  \BibitemOpen
  \bibfield  {author} {\bibinfo {author} {\bibfnamefont {A.~H.}\ \bibnamefont
  {Naghdi}}, \bibinfo {author} {\bibfnamefont {K.}~\bibnamefont {Karimi}},
  \bibinfo {author} {\bibfnamefont {A.~E.}\ \bibnamefont {Poisvert}}, \bibinfo
  {author} {\bibfnamefont {A.}~\bibnamefont {Esfandiarpour}}, \bibinfo {author}
  {\bibfnamefont {R.}~\bibnamefont {Alvarez}}, \bibinfo {author} {\bibfnamefont
  {P.}~\bibnamefont {Sobkowicz}}, \bibinfo {author} {\bibfnamefont
  {M.}~\bibnamefont {Alava}}, \ and\ \bibinfo {author} {\bibfnamefont
  {S.}~\bibnamefont {Papanikolaou}},\ }\href {\doibase
  10.1103/PhysRevB.107.094109} {\bibfield  {journal} {\bibinfo  {journal}
  {Phys. Rev. B}\ }\textbf {\bibinfo {volume} {107}},\ \bibinfo {pages}
  {094109} (\bibinfo {year} {2023})}\BibitemShut {NoStop}%
\bibitem [{\citenamefont {Karimi}\ \emph {et~al.}(2023)\citenamefont {Karimi},
  \citenamefont {Esfandiarpour},\ and\ \citenamefont
  {Papanikolaou}}]{karimi2023serrated}%
  \BibitemOpen
  \bibfield  {author} {\bibinfo {author} {\bibfnamefont {K.}~\bibnamefont
  {Karimi}}, \bibinfo {author} {\bibfnamefont {A.}~\bibnamefont
  {Esfandiarpour}}, \ and\ \bibinfo {author} {\bibfnamefont {S.}~\bibnamefont
  {Papanikolaou}},\ }\href@noop {} {\enquote {\bibinfo {title} {Serrated
  plastic flow in slowly-deforming complex concentrated alloys: universal
  signatures of dislocation avalanches},}\ } (\bibinfo {year} {2023}),\ \Eprint
  {http://arxiv.org/abs/2310.03828} {arXiv:2310.03828 [cond-mat.mtrl-sci]}
  \BibitemShut {NoStop}%
\bibitem [{\citenamefont {Karimi}\ and\ \citenamefont
  {Papanikolaou}(2023)}]{karimi2023tuning}%
  \BibitemOpen
  \bibfield  {author} {\bibinfo {author} {\bibfnamefont {K.}~\bibnamefont
  {Karimi}}\ and\ \bibinfo {author} {\bibfnamefont {S.}~\bibnamefont
  {Papanikolaou}},\ }\href@noop {} {\enquote {\bibinfo {title} {Tuning
  brittleness in multi-component metallic glasses through chemical disorder
  aging},}\ } (\bibinfo {year} {2023}),\ \Eprint
  {http://arxiv.org/abs/2309.11867} {arXiv:2309.11867 [cond-mat.soft]}
  \BibitemShut {NoStop}%
\bibitem [{\citenamefont {Behler}\ and\ \citenamefont
  {Parrinello}(2007)}]{PhysRevLett.98.146401}%
  \BibitemOpen
  \bibfield  {author} {\bibinfo {author} {\bibfnamefont {J.}~\bibnamefont
  {Behler}}\ and\ \bibinfo {author} {\bibfnamefont {M.}~\bibnamefont
  {Parrinello}},\ }\href {\doibase 10.1103/PhysRevLett.98.146401} {\bibfield
  {journal} {\bibinfo  {journal} {Phys. Rev. Lett.}\ }\textbf {\bibinfo
  {volume} {98}},\ \bibinfo {pages} {146401} (\bibinfo {year}
  {2007})}\BibitemShut {NoStop}%
\bibitem [{\citenamefont {Bart\'ok}\ \emph {et~al.}(2010)\citenamefont
  {Bart\'ok}, \citenamefont {Payne}, \citenamefont {Kondor},\ and\
  \citenamefont {Cs\'anyi}}]{PhysRevLett.104.136403}%
  \BibitemOpen
  \bibfield  {author} {\bibinfo {author} {\bibfnamefont {A.~P.}\ \bibnamefont
  {Bart\'ok}}, \bibinfo {author} {\bibfnamefont {M.~C.}\ \bibnamefont {Payne}},
  \bibinfo {author} {\bibfnamefont {R.}~\bibnamefont {Kondor}}, \ and\ \bibinfo
  {author} {\bibfnamefont {G.}~\bibnamefont {Cs\'anyi}},\ }\href {\doibase
  10.1103/PhysRevLett.104.136403} {\bibfield  {journal} {\bibinfo  {journal}
  {Phys. Rev. Lett.}\ }\textbf {\bibinfo {volume} {104}},\ \bibinfo {pages}
  {136403} (\bibinfo {year} {2010})}\BibitemShut {NoStop}%
\bibitem [{\citenamefont {Thompson}\ \emph {et~al.}(2015)\citenamefont
  {Thompson}, \citenamefont {Swiler}, \citenamefont {Trott}, \citenamefont
  {Foiles},\ and\ \citenamefont {Tucker}}]{THOMPSON2015316}%
  \BibitemOpen
  \bibfield  {author} {\bibinfo {author} {\bibfnamefont {A.}~\bibnamefont
  {Thompson}}, \bibinfo {author} {\bibfnamefont {L.}~\bibnamefont {Swiler}},
  \bibinfo {author} {\bibfnamefont {C.}~\bibnamefont {Trott}}, \bibinfo
  {author} {\bibfnamefont {S.}~\bibnamefont {Foiles}}, \ and\ \bibinfo {author}
  {\bibfnamefont {G.}~\bibnamefont {Tucker}},\ }\href {\doibase
  https://doi.org/10.1016/j.jcp.2014.12.018} {\bibfield  {journal} {\bibinfo
  {journal} {Journal of Computational Physics}\ }\textbf {\bibinfo {volume}
  {285}},\ \bibinfo {pages} {316} (\bibinfo {year} {2015})}\BibitemShut
  {NoStop}%
\bibitem [{\citenamefont {Shapeev}(2016)}]{15M1054183}%
  \BibitemOpen
  \bibfield  {author} {\bibinfo {author} {\bibfnamefont {A.~V.}\ \bibnamefont
  {Shapeev}},\ }\href {\doibase 10.1137/15M1054183} {\bibfield  {journal}
  {\bibinfo  {journal} {Multiscale Modeling \& Simulation}\ }\textbf {\bibinfo
  {volume} {14}},\ \bibinfo {pages} {1153} (\bibinfo {year} {2016})},\ \Eprint
  {http://arxiv.org/abs/https://doi.org/10.1137/15M1054183}
  {https://doi.org/10.1137/15M1054183} \BibitemShut {NoStop}%
\bibitem [{\citenamefont {Glielmo}\ \emph {et~al.}(2018)\citenamefont
  {Glielmo}, \citenamefont {Zeni},\ and\ \citenamefont
  {De~Vita}}]{PhysRevB.97.184307}%
  \BibitemOpen
  \bibfield  {author} {\bibinfo {author} {\bibfnamefont {A.}~\bibnamefont
  {Glielmo}}, \bibinfo {author} {\bibfnamefont {C.}~\bibnamefont {Zeni}}, \
  and\ \bibinfo {author} {\bibfnamefont {A.}~\bibnamefont {De~Vita}},\ }\href
  {\doibase 10.1103/PhysRevB.97.184307} {\bibfield  {journal} {\bibinfo
  {journal} {Phys. Rev. B}\ }\textbf {\bibinfo {volume} {97}},\ \bibinfo
  {pages} {184307} (\bibinfo {year} {2018})}\BibitemShut {NoStop}%
\bibitem [{\citenamefont {Zhang}\ \emph {et~al.}(2018)\citenamefont {Zhang},
  \citenamefont {Han}, \citenamefont {Wang}, \citenamefont {Car},\ and\
  \citenamefont {E}}]{PhysRevLett.120.143001}%
  \BibitemOpen
  \bibfield  {author} {\bibinfo {author} {\bibfnamefont {L.}~\bibnamefont
  {Zhang}}, \bibinfo {author} {\bibfnamefont {J.}~\bibnamefont {Han}}, \bibinfo
  {author} {\bibfnamefont {H.}~\bibnamefont {Wang}}, \bibinfo {author}
  {\bibfnamefont {R.}~\bibnamefont {Car}}, \ and\ \bibinfo {author}
  {\bibfnamefont {W.}~\bibnamefont {E}},\ }\href {\doibase
  10.1103/PhysRevLett.120.143001} {\bibfield  {journal} {\bibinfo  {journal}
  {Phys. Rev. Lett.}\ }\textbf {\bibinfo {volume} {120}},\ \bibinfo {pages}
  {143001} (\bibinfo {year} {2018})}\BibitemShut {NoStop}%
\bibitem [{\citenamefont {Behler}(2016)}]{doi:10.1063/1.4966192}%
  \BibitemOpen
  \bibfield  {author} {\bibinfo {author} {\bibfnamefont {J.}~\bibnamefont
  {Behler}},\ }\href {\doibase 10.1063/1.4966192} {\bibfield  {journal}
  {\bibinfo  {journal} {The Journal of Chemical Physics}\ }\textbf {\bibinfo
  {volume} {145}},\ \bibinfo {pages} {170901} (\bibinfo {year} {2016})},\
  \Eprint {http://arxiv.org/abs/https://doi.org/10.1063/1.4966192}
  {https://doi.org/10.1063/1.4966192} \BibitemShut {NoStop}%
\bibitem [{\citenamefont {Shaidu}\ \emph {et~al.}(2021)\citenamefont {Shaidu},
  \citenamefont {K{\"u}{\c{c}}{\"u}kbenli}, \citenamefont {Lot}, \citenamefont
  {Pellegrini}, \citenamefont {Kaxiras},\ and\ \citenamefont
  {de~Gironcoli}}]{Shaidu2021}%
  \BibitemOpen
  \bibfield  {author} {\bibinfo {author} {\bibfnamefont {Y.}~\bibnamefont
  {Shaidu}}, \bibinfo {author} {\bibfnamefont {E.}~\bibnamefont
  {K{\"u}{\c{c}}{\"u}kbenli}}, \bibinfo {author} {\bibfnamefont
  {R.}~\bibnamefont {Lot}}, \bibinfo {author} {\bibfnamefont {F.}~\bibnamefont
  {Pellegrini}}, \bibinfo {author} {\bibfnamefont {E.}~\bibnamefont {Kaxiras}},
  \ and\ \bibinfo {author} {\bibfnamefont {S.}~\bibnamefont {de~Gironcoli}},\
  }\href {\doibase 10.1038/s41524-021-00508-6} {\bibfield  {journal} {\bibinfo
  {journal} {npj Computational Materials}\ }\textbf {\bibinfo {volume} {7}},\
  \bibinfo {pages} {52} (\bibinfo {year} {2021})}\BibitemShut {NoStop}%
\bibitem [{\citenamefont {Li}\ \emph {et~al.}(2021)\citenamefont {Li},
  \citenamefont {Küçükbenli}, \citenamefont {Lam}, \citenamefont
  {Khaykovich}, \citenamefont {Kaxiras},\ and\ \citenamefont
  {Li}}]{LI2021100359}%
  \BibitemOpen
  \bibfield  {author} {\bibinfo {author} {\bibfnamefont {Q.-J.}\ \bibnamefont
  {Li}}, \bibinfo {author} {\bibfnamefont {E.}~\bibnamefont {Küçükbenli}},
  \bibinfo {author} {\bibfnamefont {S.}~\bibnamefont {Lam}}, \bibinfo {author}
  {\bibfnamefont {B.}~\bibnamefont {Khaykovich}}, \bibinfo {author}
  {\bibfnamefont {E.}~\bibnamefont {Kaxiras}}, \ and\ \bibinfo {author}
  {\bibfnamefont {J.}~\bibnamefont {Li}},\ }\href {\doibase
  https://doi.org/10.1016/j.xcrp.2021.100359} {\bibfield  {journal} {\bibinfo
  {journal} {Cell Reports Physical Science}\ }\textbf {\bibinfo {volume} {2}},\
  \bibinfo {pages} {100359} (\bibinfo {year} {2021})}\BibitemShut {NoStop}%
\bibitem [{\citenamefont {Byggm\"astar}\ \emph {et~al.}(2022)\citenamefont
  {Byggm\"astar}, \citenamefont {Nordlund},\ and\ \citenamefont
  {Djurabekova}}]{PhysRevMaterials.6.083801}%
  \BibitemOpen
  \bibfield  {author} {\bibinfo {author} {\bibfnamefont {J.}~\bibnamefont
  {Byggm\"astar}}, \bibinfo {author} {\bibfnamefont {K.}~\bibnamefont
  {Nordlund}}, \ and\ \bibinfo {author} {\bibfnamefont {F.}~\bibnamefont
  {Djurabekova}},\ }\href {\doibase 10.1103/PhysRevMaterials.6.083801}
  {\bibfield  {journal} {\bibinfo  {journal} {Phys. Rev. Mater.}\ }\textbf
  {\bibinfo {volume} {6}},\ \bibinfo {pages} {083801} (\bibinfo {year}
  {2022})}\BibitemShut {NoStop}%
\bibitem [{\citenamefont {Vandermause}\ \emph {et~al.}(2020)\citenamefont
  {Vandermause}, \citenamefont {Torrisi}, \citenamefont {Batzner},
  \citenamefont {Xie}, \citenamefont {Sun}, \citenamefont {Kolpak},\ and\
  \citenamefont {Kozinsky}}]{Vandermause2020}%
  \BibitemOpen
  \bibfield  {author} {\bibinfo {author} {\bibfnamefont {J.}~\bibnamefont
  {Vandermause}}, \bibinfo {author} {\bibfnamefont {S.~B.}\ \bibnamefont
  {Torrisi}}, \bibinfo {author} {\bibfnamefont {S.}~\bibnamefont {Batzner}},
  \bibinfo {author} {\bibfnamefont {Y.}~\bibnamefont {Xie}}, \bibinfo {author}
  {\bibfnamefont {L.}~\bibnamefont {Sun}}, \bibinfo {author} {\bibfnamefont
  {A.~M.}\ \bibnamefont {Kolpak}}, \ and\ \bibinfo {author} {\bibfnamefont
  {B.}~\bibnamefont {Kozinsky}},\ }\href {\doibase 10.1038/s41524-020-0283-z}
  {\bibfield  {journal} {\bibinfo  {journal} {npj Computational Materials}\
  }\textbf {\bibinfo {volume} {6}},\ \bibinfo {pages} {20} (\bibinfo {year}
  {2020})}\BibitemShut {NoStop}%
\bibitem [{\citenamefont {Pellegrini}\ \emph {et~al.}(2023)\citenamefont
  {Pellegrini}, \citenamefont {Lot}, \citenamefont {Shaidu},\ and\
  \citenamefont {Küçükbenli}}]{pellegrini2023panna}%
  \BibitemOpen
  \bibfield  {author} {\bibinfo {author} {\bibfnamefont {F.}~\bibnamefont
  {Pellegrini}}, \bibinfo {author} {\bibfnamefont {R.}~\bibnamefont {Lot}},
  \bibinfo {author} {\bibfnamefont {Y.}~\bibnamefont {Shaidu}}, \ and\ \bibinfo
  {author} {\bibfnamefont {E.}~\bibnamefont {Küçükbenli}},\ }\href@noop {}
  {\enquote {\bibinfo {title} {Panna 2.0: Efficient neural network interatomic
  potentials and new architectures},}\ } (\bibinfo {year} {2023}),\ \Eprint
  {http://arxiv.org/abs/2305.11805} {arXiv:2305.11805 [physics.comp-ph]}
  \BibitemShut {NoStop}%
\bibitem [{\citenamefont {Musaelian}\ \emph {et~al.}(2023)\citenamefont
  {Musaelian}, \citenamefont {Batzner}, \citenamefont {Johansson},
  \citenamefont {Sun}, \citenamefont {Owen}, \citenamefont {Kornbluth},\ and\
  \citenamefont {Kozinsky}}]{Musaelian2023}%
  \BibitemOpen
  \bibfield  {author} {\bibinfo {author} {\bibfnamefont {A.}~\bibnamefont
  {Musaelian}}, \bibinfo {author} {\bibfnamefont {S.}~\bibnamefont {Batzner}},
  \bibinfo {author} {\bibfnamefont {A.}~\bibnamefont {Johansson}}, \bibinfo
  {author} {\bibfnamefont {L.}~\bibnamefont {Sun}}, \bibinfo {author}
  {\bibfnamefont {C.~J.}\ \bibnamefont {Owen}}, \bibinfo {author}
  {\bibfnamefont {M.}~\bibnamefont {Kornbluth}}, \ and\ \bibinfo {author}
  {\bibfnamefont {B.}~\bibnamefont {Kozinsky}},\ }\href {\doibase
  10.1038/s41467-023-36329-y} {\bibfield  {journal} {\bibinfo  {journal}
  {Nature Communications}\ }\textbf {\bibinfo {volume} {14}},\ \bibinfo {pages}
  {579} (\bibinfo {year} {2023})}\BibitemShut {NoStop}%
\bibitem [{\citenamefont {Batatia}\ \emph {et~al.}(2022)\citenamefont
  {Batatia}, \citenamefont {Batzner}, \citenamefont {Kov{\'a}cs}, \citenamefont
  {Musaelian}, \citenamefont {Simm}, \citenamefont {Drautz}, \citenamefont
  {Ortner}, \citenamefont {Kozinsky},\ and\ \citenamefont
  {Cs{\'a}nyi}}]{Batatia2022Design}%
  \BibitemOpen
  \bibfield  {author} {\bibinfo {author} {\bibfnamefont {I.}~\bibnamefont
  {Batatia}}, \bibinfo {author} {\bibfnamefont {S.}~\bibnamefont {Batzner}},
  \bibinfo {author} {\bibfnamefont {D.~P.}\ \bibnamefont {Kov{\'a}cs}},
  \bibinfo {author} {\bibfnamefont {A.}~\bibnamefont {Musaelian}}, \bibinfo
  {author} {\bibfnamefont {G.~N.~C.}\ \bibnamefont {Simm}}, \bibinfo {author}
  {\bibfnamefont {R.}~\bibnamefont {Drautz}}, \bibinfo {author} {\bibfnamefont
  {C.}~\bibnamefont {Ortner}}, \bibinfo {author} {\bibfnamefont
  {B.}~\bibnamefont {Kozinsky}}, \ and\ \bibinfo {author} {\bibfnamefont
  {G.}~\bibnamefont {Cs{\'a}nyi}},\ }\href {\doibase 10.48550/arXiv.2205.06643}
  {\enquote {\bibinfo {title} {The design space of e(3)-equivariant
  atom-centered interatomic potentials},}\ } (\bibinfo {year} {2022}),\ \Eprint
  {http://arxiv.org/abs/2205.06643} {arXiv:2205.06643} \BibitemShut {NoStop}%
\bibitem [{\citenamefont {Owen}\ \emph
  {et~al.}(2023{\natexlab{a}})\citenamefont {Owen}, \citenamefont {Xie},
  \citenamefont {Johansson}, \citenamefont {Sun},\ and\ \citenamefont
  {Kozinsky}}]{owen2023stability}%
  \BibitemOpen
  \bibfield  {author} {\bibinfo {author} {\bibfnamefont {C.~J.}\ \bibnamefont
  {Owen}}, \bibinfo {author} {\bibfnamefont {Y.}~\bibnamefont {Xie}}, \bibinfo
  {author} {\bibfnamefont {A.}~\bibnamefont {Johansson}}, \bibinfo {author}
  {\bibfnamefont {L.}~\bibnamefont {Sun}}, \ and\ \bibinfo {author}
  {\bibfnamefont {B.}~\bibnamefont {Kozinsky}},\ }\href@noop {} {\enquote
  {\bibinfo {title} {Stability, mechanisms and kinetics of emergence of au
  surface reconstructions using bayesian force fields},}\ } (\bibinfo {year}
  {2023}{\natexlab{a}}),\ \Eprint {http://arxiv.org/abs/2308.07311}
  {arXiv:2308.07311 [cond-mat.mtrl-sci]} \BibitemShut {NoStop}%
\bibitem [{\citenamefont {Owen}\ \emph
  {et~al.}(2023{\natexlab{b}})\citenamefont {Owen}, \citenamefont {Marcella},
  \citenamefont {Xie}, \citenamefont {Vandermause}, \citenamefont {Frenkel},
  \citenamefont {Nuzzo},\ and\ \citenamefont {Kozinsky}}]{owen2023unraveling}%
  \BibitemOpen
  \bibfield  {author} {\bibinfo {author} {\bibfnamefont {C.~J.}\ \bibnamefont
  {Owen}}, \bibinfo {author} {\bibfnamefont {N.}~\bibnamefont {Marcella}},
  \bibinfo {author} {\bibfnamefont {Y.}~\bibnamefont {Xie}}, \bibinfo {author}
  {\bibfnamefont {J.}~\bibnamefont {Vandermause}}, \bibinfo {author}
  {\bibfnamefont {A.~I.}\ \bibnamefont {Frenkel}}, \bibinfo {author}
  {\bibfnamefont {R.~G.}\ \bibnamefont {Nuzzo}}, \ and\ \bibinfo {author}
  {\bibfnamefont {B.}~\bibnamefont {Kozinsky}},\ }\href@noop {} {\enquote
  {\bibinfo {title} {Unraveling the catalytic effect of hydrogen adsorption on
  pt nanoparticle shape-change},}\ } (\bibinfo {year} {2023}{\natexlab{b}}),\
  \Eprint {http://arxiv.org/abs/2306.00901} {arXiv:2306.00901
  [cond-mat.mtrl-sci]} \BibitemShut {NoStop}%
\bibitem [{\citenamefont {Goryaeva}\ \emph {et~al.}(2021)\citenamefont
  {Goryaeva}, \citenamefont {D\'er\`es}, \citenamefont {Lapointe},
  \citenamefont {Grigorev}, \citenamefont {Swinburne}, \citenamefont {Kermode},
  \citenamefont {Ventelon}, \citenamefont {Baima},\ and\ \citenamefont
  {Marinica}}]{PhysRevMaterials.5.103803}%
  \BibitemOpen
  \bibfield  {author} {\bibinfo {author} {\bibfnamefont {A.~M.}\ \bibnamefont
  {Goryaeva}}, \bibinfo {author} {\bibfnamefont {J.}~\bibnamefont {D\'er\`es}},
  \bibinfo {author} {\bibfnamefont {C.}~\bibnamefont {Lapointe}}, \bibinfo
  {author} {\bibfnamefont {P.}~\bibnamefont {Grigorev}}, \bibinfo {author}
  {\bibfnamefont {T.~D.}\ \bibnamefont {Swinburne}}, \bibinfo {author}
  {\bibfnamefont {J.~R.}\ \bibnamefont {Kermode}}, \bibinfo {author}
  {\bibfnamefont {L.}~\bibnamefont {Ventelon}}, \bibinfo {author}
  {\bibfnamefont {J.}~\bibnamefont {Baima}}, \ and\ \bibinfo {author}
  {\bibfnamefont {M.-C.}\ \bibnamefont {Marinica}},\ }\href {\doibase
  10.1103/PhysRevMaterials.5.103803} {\bibfield  {journal} {\bibinfo  {journal}
  {Phys. Rev. Mater.}\ }\textbf {\bibinfo {volume} {5}},\ \bibinfo {pages}
  {103803} (\bibinfo {year} {2021})}\BibitemShut {NoStop}%
\bibitem [{\citenamefont {Byggm\"astar}\ \emph {et~al.}(2019)\citenamefont
  {Byggm\"astar}, \citenamefont {Hamedani}, \citenamefont {Nordlund},\ and\
  \citenamefont {Djurabekova}}]{PhysRevB.100.144105}%
  \BibitemOpen
  \bibfield  {author} {\bibinfo {author} {\bibfnamefont {J.}~\bibnamefont
  {Byggm\"astar}}, \bibinfo {author} {\bibfnamefont {A.}~\bibnamefont
  {Hamedani}}, \bibinfo {author} {\bibfnamefont {K.}~\bibnamefont {Nordlund}},
  \ and\ \bibinfo {author} {\bibfnamefont {F.}~\bibnamefont {Djurabekova}},\
  }\href {\doibase 10.1103/PhysRevB.100.144105} {\bibfield  {journal} {\bibinfo
   {journal} {Phys. Rev. B}\ }\textbf {\bibinfo {volume} {100}},\ \bibinfo
  {pages} {144105} (\bibinfo {year} {2019})}\BibitemShut {NoStop}%
\bibitem [{\citenamefont {Nikoulis}\ \emph {et~al.}(2021)\citenamefont
  {Nikoulis}, \citenamefont {Byggmästar}, \citenamefont {Kioseoglou},
  \citenamefont {Nordlund},\ and\ \citenamefont {Djurabekova}}]{Nikoulis_2021}%
  \BibitemOpen
  \bibfield  {author} {\bibinfo {author} {\bibfnamefont {G.}~\bibnamefont
  {Nikoulis}}, \bibinfo {author} {\bibfnamefont {J.}~\bibnamefont
  {Byggmästar}}, \bibinfo {author} {\bibfnamefont {J.}~\bibnamefont
  {Kioseoglou}}, \bibinfo {author} {\bibfnamefont {K.}~\bibnamefont
  {Nordlund}}, \ and\ \bibinfo {author} {\bibfnamefont {F.}~\bibnamefont
  {Djurabekova}},\ }\href {\doibase 10.1088/1361-648X/ac03d1} {\bibfield
  {journal} {\bibinfo  {journal} {Journal of Physics: Condensed Matter}\
  }\textbf {\bibinfo {volume} {33}},\ \bibinfo {pages} {315403} (\bibinfo
  {year} {2021})}\BibitemShut {NoStop}%
\bibitem [{\citenamefont {Byggm\"astar}\ \emph {et~al.}(2021)\citenamefont
  {Byggm\"astar}, \citenamefont {Nordlund},\ and\ \citenamefont
  {Djurabekova}}]{PhysRevB.104.104101}%
  \BibitemOpen
  \bibfield  {author} {\bibinfo {author} {\bibfnamefont {J.}~\bibnamefont
  {Byggm\"astar}}, \bibinfo {author} {\bibfnamefont {K.}~\bibnamefont
  {Nordlund}}, \ and\ \bibinfo {author} {\bibfnamefont {F.}~\bibnamefont
  {Djurabekova}},\ }\href {\doibase 10.1103/PhysRevB.104.104101} {\bibfield
  {journal} {\bibinfo  {journal} {Phys. Rev. B}\ }\textbf {\bibinfo {volume}
  {104}},\ \bibinfo {pages} {104101} (\bibinfo {year} {2021})}\BibitemShut
  {NoStop}%
\bibitem [{\citenamefont {Byggm\"astar}\ \emph
  {et~al.}(2020{\natexlab{a}})\citenamefont {Byggm\"astar}, \citenamefont
  {Nordlund},\ and\ \citenamefont {Djurabekova}}]{dataset}%
  \BibitemOpen
  \bibfield  {author} {\bibinfo {author} {\bibfnamefont {J.}~\bibnamefont
  {Byggm\"astar}}, \bibinfo {author} {\bibfnamefont {K.}~\bibnamefont
  {Nordlund}}, \ and\ \bibinfo {author} {\bibfnamefont {F.}~\bibnamefont
  {Djurabekova}},\ }\href {\doibase 10.1103/PhysRevMaterials.4.093802}
  {\bibfield  {journal} {\bibinfo  {journal} {Phys. Rev. Materials}\ }\textbf
  {\bibinfo {volume} {4}},\ \bibinfo {pages} {093802} (\bibinfo {year}
  {2020}{\natexlab{a}})}\BibitemShut {NoStop}%
\bibitem [{\citenamefont {Lot}\ \emph {et~al.}(2020)\citenamefont {Lot},
  \citenamefont {Pellegrini}, \citenamefont {Shaidu},\ and\ \citenamefont
  {Küçükbenli}}]{panna}%
  \BibitemOpen
  \bibfield  {author} {\bibinfo {author} {\bibfnamefont {R.}~\bibnamefont
  {Lot}}, \bibinfo {author} {\bibfnamefont {F.}~\bibnamefont {Pellegrini}},
  \bibinfo {author} {\bibfnamefont {Y.}~\bibnamefont {Shaidu}}, \ and\ \bibinfo
  {author} {\bibfnamefont {E.}~\bibnamefont {Küçükbenli}},\ }\href {\doibase
  https://doi.org/10.1016/j.cpc.2020.107402} {\bibfield  {journal} {\bibinfo
  {journal} {Computer Physics Communications}\ }\textbf {\bibinfo {volume}
  {256}},\ \bibinfo {pages} {107402} (\bibinfo {year} {2020})}\BibitemShut
  {NoStop}%
\bibitem [{\citenamefont {Salonen}\ \emph {et~al.}(2003)\citenamefont
  {Salonen}, \citenamefont {JĂ¤rvi}, \citenamefont {Nordlund},\ and\
  \citenamefont {Keinonen}}]{ESalonen_2003}%
  \BibitemOpen
  \bibfield  {author} {\bibinfo {author} {\bibfnamefont {E.}~\bibnamefont
  {Salonen}}, \bibinfo {author} {\bibfnamefont {T.}~\bibnamefont {JĂ¤rvi}},
  \bibinfo {author} {\bibfnamefont {K.}~\bibnamefont {Nordlund}}, \ and\
  \bibinfo {author} {\bibfnamefont {J.}~\bibnamefont {Keinonen}},\ }\href
  {\doibase 10.1088/0953-8984/15/34/314} {\bibfield  {journal} {\bibinfo
  {journal} {Journal of Physics: Condensed Matter}\ }\textbf {\bibinfo {volume}
  {15}},\ \bibinfo {pages} {5845} (\bibinfo {year} {2003})}\BibitemShut
  {NoStop}%
\bibitem [{\citenamefont {Smith}\ \emph {et~al.}(2017)\citenamefont {Smith},
  \citenamefont {Isayev},\ and\ \citenamefont {Roitberg}}]{C6SC05720A}%
  \BibitemOpen
  \bibfield  {author} {\bibinfo {author} {\bibfnamefont {J.~S.}\ \bibnamefont
  {Smith}}, \bibinfo {author} {\bibfnamefont {O.}~\bibnamefont {Isayev}}, \
  and\ \bibinfo {author} {\bibfnamefont {A.~E.}\ \bibnamefont {Roitberg}},\
  }\href {\doibase 10.1039/C6SC05720A} {\bibfield  {journal} {\bibinfo
  {journal} {Chem. Sci.}\ }\textbf {\bibinfo {volume} {8}},\ \bibinfo {pages}
  {3192} (\bibinfo {year} {2017})}\BibitemShut {NoStop}%
\bibitem [{\citenamefont {Rumble}(2019)}]{rumble2019crc}%
  \BibitemOpen
  \bibfield  {author} {\bibinfo {author} {\bibfnamefont {J.}~\bibnamefont
  {Rumble}},\ }\href@noop {} {\emph {\bibinfo {title} {CRC Handbook of
  Chemistry and Physics, 100th Edition}}},\ \bibinfo {edition} {100th}\ ed.\
  (\bibinfo  {publisher} {CRC Press},\ \bibinfo {address} {Boca Raton, FL},\
  \bibinfo {year} {2019})\BibitemShut {NoStop}%
\bibitem [{\citenamefont {M{\"o}ller}\ and\ \citenamefont
  {Bitzek}(2016)}]{moller2016bda}%
  \BibitemOpen
  \bibfield  {author} {\bibinfo {author} {\bibfnamefont {J.~J.}\ \bibnamefont
  {M{\"o}ller}}\ and\ \bibinfo {author} {\bibfnamefont {E.}~\bibnamefont
  {Bitzek}},\ }\href {\doibase 10.1016/j.mex.2016.03.013} {\bibfield  {journal}
  {\bibinfo  {journal} {MethodsX}\ }\textbf {\bibinfo {volume} {3}},\ \bibinfo
  {pages} {279} (\bibinfo {year} {2016})}\BibitemShut {NoStop}%
\bibitem [{\citenamefont {Mulewska}\ \emph {et~al.}(2023)\citenamefont
  {Mulewska}, \citenamefont {Dominguez-Gutierrez}, \citenamefont {Kalita},
  \citenamefont {ByggmAstar}, \citenamefont {Wei}, \citenamefont
  {ChromiĹ„ski}, \citenamefont {Papanikolaou}, \citenamefont {Alava},
  \citenamefont {Kurpaska},\ and\ \citenamefont
  {Jagielski}}]{MULEWSKA2023154690}%
  \BibitemOpen
  \bibfield  {author} {\bibinfo {author} {\bibfnamefont {K.}~\bibnamefont
  {Mulewska}}, \bibinfo {author} {\bibfnamefont {F.}~\bibnamefont
  {Dominguez-Gutierrez}}, \bibinfo {author} {\bibfnamefont {D.}~\bibnamefont
  {Kalita}}, \bibinfo {author} {\bibfnamefont {J.}~\bibnamefont {ByggmAstar}},
  \bibinfo {author} {\bibfnamefont {G.}~\bibnamefont {Wei}}, \bibinfo {author}
  {\bibfnamefont {W.}~\bibnamefont {ChromiĹ„ski}}, \bibinfo {author}
  {\bibfnamefont {S.}~\bibnamefont {Papanikolaou}}, \bibinfo {author}
  {\bibfnamefont {M.}~\bibnamefont {Alava}}, \bibinfo {author} {\bibfnamefont
  {L.}~\bibnamefont {Kurpaska}}, \ and\ \bibinfo {author} {\bibfnamefont
  {J.}~\bibnamefont {Jagielski}},\ }\href {\doibase
  https://doi.org/10.1016/j.jnucmat.2023.154690} {\bibfield  {journal}
  {\bibinfo  {journal} {Journal of Nuclear Materials}\ }\textbf {\bibinfo
  {volume} {586}},\ \bibinfo {pages} {154690} (\bibinfo {year}
  {2023})}\BibitemShut {NoStop}%
\bibitem [{\citenamefont {et~al.}(2015)}]{tensorflow2015-whitepaper}%
  \BibitemOpen
  \bibfield  {author} {\bibinfo {author} {\bibfnamefont {M.~A.}\ \bibnamefont
  {et~al.}},\ }\href {https://www.tensorflow.org/} {\enquote {\bibinfo {title}
  {{TensorFlow}: Large-scale machine learning on heterogeneous systems},}\ }
  (\bibinfo {year} {2015}),\ \bibinfo {note} {software available from
  tensorflow.org}\BibitemShut {NoStop}%
\bibitem [{\citenamefont {Byggm\"astar}\ \emph
  {et~al.}(2020{\natexlab{b}})\citenamefont {Byggm\"astar}, \citenamefont
  {Nordlund},\ and\ \citenamefont {Djurabekova}}]{PhysRevMaterials.4.093802}%
  \BibitemOpen
  \bibfield  {author} {\bibinfo {author} {\bibfnamefont {J.}~\bibnamefont
  {Byggm\"astar}}, \bibinfo {author} {\bibfnamefont {K.}~\bibnamefont
  {Nordlund}}, \ and\ \bibinfo {author} {\bibfnamefont {F.}~\bibnamefont
  {Djurabekova}},\ }\href {\doibase 10.1103/PhysRevMaterials.4.093802}
  {\bibfield  {journal} {\bibinfo  {journal} {Phys. Rev. Mater.}\ }\textbf
  {\bibinfo {volume} {4}},\ \bibinfo {pages} {093802} (\bibinfo {year}
  {2020}{\natexlab{b}})}\BibitemShut {NoStop}%
\bibitem [{\citenamefont {et~al.}(2009)}]{Giannozzi_2009}%
  \BibitemOpen
  \bibfield  {author} {\bibinfo {author} {\bibfnamefont {P.~G.}\ \bibnamefont
  {et~al.}},\ }\href {\doibase 10.1088/0953-8984/21/39/395502} {\bibfield
  {journal} {\bibinfo  {journal} {Journal of Physics: Condensed Matter}\
  }\textbf {\bibinfo {volume} {21}},\ \bibinfo {pages} {395502} (\bibinfo
  {year} {2009})}\BibitemShut {NoStop}%
\bibitem [{\citenamefont {et~al.}(2017)}]{Giannozzi_2017}%
  \BibitemOpen
  \bibfield  {author} {\bibinfo {author} {\bibfnamefont {P.~G.}\ \bibnamefont
  {et~al.}},\ }\href {\doibase 10.1088/1361-648x/aa8f79} {\bibfield  {journal}
  {\bibinfo  {journal} {Journal of Physics: Condensed Matter}\ }\textbf
  {\bibinfo {volume} {29}},\ \bibinfo {pages} {465901} (\bibinfo {year}
  {2017})}\BibitemShut {NoStop}%
\bibitem [{\citenamefont {Prandini}\ \emph {et~al.}(2018)\citenamefont
  {Prandini}, \citenamefont {Marrazzo}, \citenamefont {Castelli}, \citenamefont
  {Mounet},\ and\ \citenamefont {Marzari}}]{Prandini2018}%
  \BibitemOpen
  \bibfield  {author} {\bibinfo {author} {\bibfnamefont {G.}~\bibnamefont
  {Prandini}}, \bibinfo {author} {\bibfnamefont {A.}~\bibnamefont {Marrazzo}},
  \bibinfo {author} {\bibfnamefont {I.~E.}\ \bibnamefont {Castelli}}, \bibinfo
  {author} {\bibfnamefont {N.}~\bibnamefont {Mounet}}, \ and\ \bibinfo {author}
  {\bibfnamefont {N.}~\bibnamefont {Marzari}},\ }\href {\doibase
  10.1038/s41524-018-0127-2} {\bibfield  {journal} {\bibinfo  {journal} {npj
  Computational Materials}\ }\textbf {\bibinfo {volume} {4}},\ \bibinfo {pages}
  {72} (\bibinfo {year} {2018})}\BibitemShut {NoStop}%
\bibitem [{\citenamefont {et~al.}(2016)}]{doi:10.1126/science.aad3000}%
  \BibitemOpen
  \bibfield  {author} {\bibinfo {author} {\bibfnamefont {K.~L.}\ \bibnamefont
  {et~al.}},\ }\href {\doibase 10.1126/science.aad3000} {\bibfield  {journal}
  {\bibinfo  {journal} {Science}\ }\textbf {\bibinfo {volume} {351}},\ \bibinfo
  {pages} {aad3000} (\bibinfo {year} {2016})},\ \Eprint
  {http://arxiv.org/abs/https://www.science.org/doi/pdf/10.1126/science.aad3000}
  {https://www.science.org/doi/pdf/10.1126/science.aad3000} \BibitemShut
  {NoStop}%
\bibitem [{\citenamefont {Hamann}(2013)}]{PhysRevB.88.085117}%
  \BibitemOpen
  \bibfield  {author} {\bibinfo {author} {\bibfnamefont {D.~R.}\ \bibnamefont
  {Hamann}},\ }\href {\doibase 10.1103/PhysRevB.88.085117} {\bibfield
  {journal} {\bibinfo  {journal} {Phys. Rev. B}\ }\textbf {\bibinfo {volume}
  {88}},\ \bibinfo {pages} {085117} (\bibinfo {year} {2013})}\BibitemShut
  {NoStop}%
\bibitem [{\citenamefont {Monkhorst}\ and\ \citenamefont
  {Pack}(1976)}]{PhysRevB.13.5188}%
  \BibitemOpen
  \bibfield  {author} {\bibinfo {author} {\bibfnamefont {H.~J.}\ \bibnamefont
  {Monkhorst}}\ and\ \bibinfo {author} {\bibfnamefont {J.~D.}\ \bibnamefont
  {Pack}},\ }\href {\doibase 10.1103/PhysRevB.13.5188} {\bibfield  {journal}
  {\bibinfo  {journal} {Phys. Rev. B}\ }\textbf {\bibinfo {volume} {13}},\
  \bibinfo {pages} {5188} (\bibinfo {year} {1976})}\BibitemShut {NoStop}%
\bibitem [{\citenamefont {Methfessel}\ and\ \citenamefont
  {Paxton}(1989)}]{PhysRevB.40.3616}%
  \BibitemOpen
  \bibfield  {author} {\bibinfo {author} {\bibfnamefont {M.}~\bibnamefont
  {Methfessel}}\ and\ \bibinfo {author} {\bibfnamefont {A.~T.}\ \bibnamefont
  {Paxton}},\ }\href {\doibase 10.1103/PhysRevB.40.3616} {\bibfield  {journal}
  {\bibinfo  {journal} {Phys. Rev. B}\ }\textbf {\bibinfo {volume} {40}},\
  \bibinfo {pages} {3616} (\bibinfo {year} {1989})}\BibitemShut {NoStop}%
\bibitem [{\citenamefont {Chung}\ and\ \citenamefont
  {Buessem}(1967)}]{chung1967voigt}%
  \BibitemOpen
  \bibfield  {author} {\bibinfo {author} {\bibfnamefont {D.}~\bibnamefont
  {Chung}}\ and\ \bibinfo {author} {\bibfnamefont {W.}~\bibnamefont
  {Buessem}},\ }\href@noop {} {\bibfield  {journal} {\bibinfo  {journal}
  {Journal of Applied Physics}\ }\textbf {\bibinfo {volume} {38}},\ \bibinfo
  {pages} {2535} (\bibinfo {year} {1967})}\BibitemShut {NoStop}%
\bibitem [{the()}]{thermopw}%
  \BibitemOpen
  \href {https://dalcorso.github.io/thermo\_pw/} {\enquote {\bibinfo {title}
  {Thermo\_pw driver for qe (github link).}}\ }\BibitemShut {NoStop}%
\bibitem [{\citenamefont {Kingma}\ and\ \citenamefont
  {Ba}(2017)}]{kingma2017adam}%
  \BibitemOpen
  \bibfield  {author} {\bibinfo {author} {\bibfnamefont {D.~P.}\ \bibnamefont
  {Kingma}}\ and\ \bibinfo {author} {\bibfnamefont {J.}~\bibnamefont {Ba}},\
  }\href@noop {} {\enquote {\bibinfo {title} {Adam: A method for stochastic
  optimization},}\ } (\bibinfo {year} {2017}),\ \Eprint
  {http://arxiv.org/abs/1412.6980} {arXiv:1412.6980 [cs.LG]} \BibitemShut
  {NoStop}%
\end{thebibliography}%
